%% file: paper.tex
\documentclass[%
dvipsnames,
reprint,
superscriptaddress,
 prb,
]{revtex4-1}

\usepackage[T1]{fontenc}             
\usepackage[utf8]{inputenc}          
\usepackage[english]{babel}          

\usepackage{amsmath}                 
\usepackage{amssymb}                 
\usepackage{mathrsfs}                
\usepackage{mathtools}               
\usepackage{dsfont}                  

\usepackage{graphicx}
\usepackage{bm}
\usepackage{microtype}                 

\usepackage{standalone}
\usepackage{tikz}
\usepackage{xcolor}
\usetikzlibrary{backgrounds,fadings}
\usetikzlibrary{calc}
\usetikzlibrary{arrows}
\usetikzlibrary{decorations.pathreplacing}

\usepackage{siunitx}                   
\usepackage{physics}
\renewcommand*{\Im}{\imaginary}
\renewcommand*{\Re}{\real}
\usepackage[caption=false]{subfig}

\usepackage[unicode=true]{hyperref}
\hypersetup{%
  pdftitle={Spectral properties of half-metallic ferromagnets},
  pdfsubject={half-metallic ferromagnets, local many-body correlation, DMFT},
  pdfauthor={A. Weh, J. Otsuki, H. Schnait, H. G. Evertz, U. Eckern, A. I. Lichtenstein, L. Chioncel},
  pdfkeywords={HMF, DMFT, CT-QMC, DMRG}
}
\usepackage[all]{hypcap} 
\usepackage[capitalise]{cleveref}
\crefname{section}{Sec.}{Sections}
\Crefname{section}{Section}{Sections}
\newcommand{\eschot}{e^\text{s}}
\input{commands}
\input{tikz/captionize.tikz.tex}

\begin{document}

\title{Spectral properties of heterostructures containing half-metallic ferromagnets in the presence of local many-body correlations}

\author{A. Weh}
\email{andreas.weh@physik.uni-augsburg.de}
\affiliation{%
  Theoretical Physics II, Institute of Physics, University of Augsburg, 86135 Augsburg, Germany
}
\author{J. Otsuki}%
\affiliation{%
  Research Institute for Interdisciplinary Science, Okayama University, Okayama 700--8530, Japan
}%
\author{H. Schnait}%
\affiliation{Institute of Theoretical and Computational Physics, Graz University of Technology, 8010 Graz, Austria}
\author{H. G. Evertz}%
\affiliation{Institute of Theoretical and Computational Physics, Graz University of Technology, 8010 Graz, Austria}
\author{U. Eckern}
\affiliation{%
  Theoretical Physics II, Institute of Physics, University of Augsburg, 86135 Augsburg, Germany
}
\author{A. I. Lichtenstein}
\affiliation{%
  Institute of Theoretical Physics, University of Hamburg, Jungiusstraße 9, 20355 Hamburg, Germany}
\affiliation{%
  The Hamburg Centre for Ultrafast Imaging, Luruper Chaussee 149, 22761 Hamburg, Germany
}
\author{L. Chioncel}
\affiliation{%
  Augsburg Center for Innovative Technologies, University of Augsburg, 86135 Augsburg, Germany
}%
\affiliation{%
  Theoretical Physics III, Center for Electronic Correlations and Magnetism,
  Institute of Physics, University of Augsburg, 86135 Augsburg, Germany
}%

\date{\today}

\begin{abstract}
In this work, we investigate models for bulk, bi- and multilayers containing 
half-metallic ferromagnets (HMFs), at zero and at finite temperature, in order 
to elucidate the effects of strong electronic correlations on the spectral 
properties (density of states).
Our focus is on the evolution of the finite-temperature many-body induced tails 
in the half-metallic gap.
To this end, the dynamical mean-field theory (DMFT) is employed.
For the bulk, a Bethe lattice model is solved using
a matrix product states based impurity solver at zero temperature
and a continuous-time quantum Monte Carlo (CT-QMC) solver at finite temperature.
We demonstrate numerically, in agreement with the analytical result,
that the tails vanish at the Fermi level at zero temperature.
In order to study multilayers, taken to be square lattices within the layers,
we use the real-space DMFT extension with the CT-QMC impurity solver.
For bilayers formed by the HMF with a band or correlated insulator,
we find that charge fluctuations between the layers enhance the finite 
temperature tails.
In addition, in the presence of inter-layer hopping,
a coherent quasiparticle peak forms in the otherwise correlated insulator.
In the multilayer heterostructure setup,
we find that by suitably choosing the model parameters,
the tails at the HMF/Mott insulator interface can be reduced significantly,
and that a high spin polarization is conceivable,
even in the presence of long-ranged electrostatic interactions.
\end{abstract}

\maketitle

\section{Introduction}

A half-metal is a material that has a metallic density of states at the Fermi level 
for one spin channel and simultaneously a band gap for the other spin channel.
This extreme asymmetry between the spin channels is the source 
of great promise for spintronic application\cite{zu.fa.04,katsnelson_halfmetallic_2008}.
Half-metallic electrodes could provide fully spin-polarized 
currents and large magnetoresistance in giant magnetoresistance 
and tunnel magnetoresistance 
devices\cite{katsnelson_halfmetallic_2008}.
Density-functional theory (DFT)\cite{ho.ko.64,kohn.99,jo.gu.89,jone.15} studies have identified a number of half-metallic bulk compounds, 
including Heusler alloys\cite{degroot_new_1983}, double perovskites, transition metal oxides, chalcogenides, and pnictides.
Some of these proposed materials have been realized in experiments.
Within DFT the ground states of these materials 
are accessible from a single particle picture. 
However, whenever many-body
effects are essential the band theory is expected to fail\cite{oka_interfaces_2005}.
In particular, in metallic ferromagnets spin fluctuations play a crucial role\cite{moriya2012}.
Therefore, the scattering of charge carriers on such magnetic excitations is expected to 
influence the macroscopic properties of these materials including transport.

Contrary to the itinerant ferromagnets in which states near the Fermi level are 
quasiparticles for both spin projections,
in half-metallic ferromagnets (HMFs) an important role is played by incoherent 
non-quasiparticle (NQP) states.
These occur near the Fermi level in the energy gap\cite{Edwards_1973,ir.ka.83,ir.ka.90,ir.ka.94},
and their tails crosses the Fermi level and produce significant
depolarization effects\cite{katsnelson_halfmetallic_2008}.
The density of the NQP states vanishes at the Fermi level
but increases strongly on an energy scale of the order of the 
characteristic magnon frequency. 
Note the clear distinction between the minority and majority half-metallic cases,
corresponding to almost empty and almost full occupation of the insulating spin channel.
While for the minority gap HMF the NQP states are located just above the Fermi level,
for the majority gap HMF they appear just below the Fermi level.
The NQP states 
are expected to contribute to several physical properties
such as polarization, specific heat, and transport\cite{katsnelson_halfmetallic_2008,we.qi.19}.
In the limit of very strong interactions and close to a completely polarized band,
a significant logarithmic singularity exists in the imaginary 
part of the Green’s function, which corresponds to a 
finite jump in the density of states\cite{ir.ka.85}.
Edwards and Herz investigated the stability of the 
saturated ferromagnetic state using Green's function
methods, which agree in the limit of large interactions
with variational calculations\cite{li.ed.91}. 

It should be noted that dynamical mean-field theory (DMFT) yields qualitatively similar 
results in the limit of large $U/t$\cite{ob.pr.97}: the saturated 
ferromagnetic state is stable, however for realistic
values of $U$ its stability is far from being obvious. (As usual, $U$ and $t$ denote the local
interaction and the hopping amplitude, respectively.)
DMFT\cite{metzner_correlated_1989,georges_hubbard_1992,georges_dynamical_1996,kotliar_strongly_2004}
in combination with first principles\cite{held.07,ko.sa.06} calculations 
have been performed for the prototype HMF,
NiMnSb\cite{ch.ka.03}, as
well as for other Heusler alloys\cite{ch.ar.06,ch.ar.09}, zinc-blende structure
compounds\cite{ch.ka.05,ch.ma.06}, and CrO$_2$\cite{ch.al.07}. Using cluster 
approaches beyond the local DMFT, the many-body
features were found to be enhanced\cite{al.ch.10,mo.al.12}.
While these effects have been studied for bulk half-metallic systems, less is known for heterostructures containing half-metals except some specific cases of zinc-blende
structures\cite{ch.le.11,be.ho.11}.

In this paper, we describe the behavior of the many-body induced tails in the 
half-metallic gap for bulk systems as well as for multilayers using model studies.
In our approach we consider bi- and multilayers consisting of 
a finite number of half-metallic layers in contact with 
different numbers of metallic or insulating layers.
It is expected that away from the interface
half-metallicity is preserved on the HMF side.
At the interface region parameter optimization is important to preserve
half-metallicity.
This involves the control of electronic states in the vicinity of the 
gap to increase the spin-polarization (i.e., reduce the
interaction-induced tails) at finite temperatures.
In order to produce a half-metal in the interface, a band-gap at the Fermi level
either in the minority- or in the majority-spin spectral function 
(density of states) needs to be created.

Using different single-band Hubbard Hamiltonians on distinct 
layers allows for optimization of their
parameters such as the magnitude of the inter-layer hoppings, 
strength of local interactions, 
on-site energies, and Zeeman splittings. 
Therefore, we study such Hubbard Hamiltonians\cite{hubbard_electron_1963}
for the multilayer using the DMFT and its real space extension (R-DMFT)\cite{potthoff_metallic_1999,freericks_dynamical_2004}.
R-DMFT considers a purely local self-energy for the strong electron correlation in the layers.
To study electronic charge reconstruction,
we extend the Hubbard Hamiltonian to include long-ranged Coulomb repulsion between the layers.
We treat the latter on a mean-field level, calculating the
electrostatic potential self-consistently from the Poisson equation. 
For a multilayer of five HMFs and the same number
of Mott insulator layers, the effect of the long-ranged
repulsion is found to lead to a slight redistribution of charges
in the metallic channel.

The focus of our analysis is on a 
narrow energy range around the Fermi level.
We show that by analytic continuation of the self-energy
(instead of the Green's function)
to real energies,
robust numerical results can be obtained. 
Some preliminary results have been
reported recently\cite{we.ot.19}.
We demonstrate that 
many-body effects (described by DMFT) lead to a dynamical reduction 
of the Hartree part of the self-energy.
Therefore, the splitting between majority and minority spin channels 
is reduced. Furthermore, a temperature dependent tail emerges in the
half-metallic gap, reducing the polarization at high temperatures.
The magnitude of these effects can be 
modified by the optimization of Hamiltonian 
parameters. We expect that our results will be useful for 
a systematic engineering of heterostructures containing half-metals
with desired properties.

The paper is organized as follows. After the introductory section
we provide in \cref{sec:method} the computational details, and discuss 
the relevant parameters and methods used to solve the 
bulk system as well as the multilayer setup.
For completeness, we have included the derivation of the R-DMFT equations in \cref{sec:RDMFT-derivation}.
The results section, \cref{sec:results}, starts with a discussion of the finite 
temperature behavior of the half-metallic gap, in particular,
of the results for the spectral function and the susceptibility,
and compares them with previous calculations.
The analytic continuations of the self-energy and the Green's function
for the finite temperature spectral functions are compared in \cref{sec:pade}.
This is followed by the results for the bilayer structure, \cref{sec:bilayer},
where a square lattice is considered within the layers. 
In \cref{sec:multilayers} we consider a heterostructure of 
five half-metallic and five Mott insulator layers.
Finally, \cref{sec:conclusion} presents the conclusions of our work.

\section{Computational Method}\label{sec:method}
We use a single-band Hubbard model to describe correlation 
effects in bulk, bi- and multilayer systems.
The system Hamiltonian reads:
\begin{equation}\label{eq:Hubbard-Hamiltonian}
  \hat{H} 
  = \sum_{i, \sigma}  \tilde{\epsilon}_{i\sigma} \hat{n}_{i \sigma}
  + \sum_{ij, \sigma} {t}_{ij} \ocd_{i\sigma} \oc_{j\sigma}
  +\sum_{i} U_i \hat{n}_{i\uparrow} \hat{n}_{i\downarrow}.
\end{equation}
Here \(\ocd_{i\sigma}\) and \(\oc_{i\sigma}\) are the fermionic creation and annihilation
operators at site \(i\) with spin \(\sigma\).
We denote the number operator at site \(i\) with \(\hat{n}_{i\sigma} = \ocd_{i\sigma} \oc_{i\sigma}\).
The on-site contributions are \(\tilde{\epsilon}_{i \sigma} = \epsilon_{i} + \sigma h_{i} - \mu - U_{i}/2\),
with the on-site energy \(\epsilon_{i}\),
the magnetic splitting \(h_{i}\),
and the chemical potential \(\mu\).
Furthermore,
the parameters \({t}_{ij}\) are the hopping matrix elements,
and \(U_{i}\) is the Hubbard interaction.
The hopping matrix is Hermitian, \(t_{ij} = t_{ji}^*\).
While there exists a solution for a one-dimensional system\cite{h.l.essler_onedimensional_2010},
we cannot solve this problem in general in higher dimensions.
The dynamical mean-field theory (DMFT)\cite{metzner_correlated_1989,georges_hubbard_1992,georges_dynamical_1996,kotliar_strongly_2004},
however, provides a non-perturbative approach
which is applicable for any range of parameters, and is exact in the limit of infinite coordination number.
It is furthermore exact for both solvable limits,
the non-interacting case, \(U_i=0\), and the atomic limit, \(t_{ij}=0\).
Continuous-time quantum Monte Carlo (CT-QMC) methods are a tool of choice to 
solve correlated electron problems\cite{gull_continuoustime_2011}. In the context 
of DMFT, the Hubbard model in the limit of infinite coordination number maps onto 
that of the single-impurity Anderson model (SIAM),
which leads to invaluable insight into the Mott transition~\cite{georges_dynamical_1996}.
Being an action-based method,
it allows the simulation of effective low-energy models 
after integrating out high-energy degrees of freedom. 
Further applications of the CT-QMC include, e.g., formulations along the 
Keldysh contour, applications within the cluster extensions of DMFT to include 
spatial fluctuations\cite{maier_quantum_2005},
the dual fermion approach\cite{ru.ka.08},
or the dynamical vertex approximation\cite{to.ka.07}.
The CT-QMC methods have different formulations:
the interaction expansion (CT-INT)\cite{ru.sa.05}, the 
auxiliary-field (CT-AUX)\cite{gu.we.08}, and the hybridization 
expansion (CT-HYB)\cite{we.co.06}.
We use the CT-HYB formulation for all finite temperature results presented here,
since the efficiency of the segment picture was shown for single-band problems\cite{gull_continuoustime_2011}.
CT-QMC operates on ``imaginary time'', therefore an analytical continuation is 
necessary to produce spectral functions on the real-frequency axis.
This is an ill-conditioned problem and limits the precision of calculated spectral functions.
This issue may be especially severe for
multi-orbital problems with complicated spectral lines.
A solution to this problem is to do time evolution directly on the real time axis
utilizing matrix product state (MPS) based impurity solvers\cite{Wang_Zhuang_Dai_Xie_2010,Wolf_McCulloch_Schollwock_2014,ganahl2015efficient,bauernfeind2019comparison}.
These solvers allow for a precise discretization of the hybridization function (with several hundred bath sites per spin)
and have been shown to yield excellent results even for sharp peaks at high energies in the spectral function\cite{ganahl2015efficient}.
They have been generalized to multi-orbital impurity solvers\cite{Bauernfeind_Zingl_Triebl_Aichhorn_Evertz_2017,Bauernfeind_Triebl_Zingl_Aichhorn_Evertz_2018,Bauernfeind_Aichhorn_2020}
by employing tensor-network representations of the impurity model,
while still keeping very good results at all energies, with moderate computational cost.
For the present work, we have extended the MPS solver presented in Refs.\ \onlinecite{Bauernfeind_Zingl_Triebl_Aichhorn_Evertz_2017,bauernfeind2019comparison}.
to allow for magnetically polarized calculations at zero temperature, \(T=0\).
We calculate the ground state using the density matrix renormalization group (DMRG)\cite{White_1992,schollwock2011density}
and we perform the time evolution using time-evolving block decimation (TEBD)\cite{Vidal_2003,Vidal_2004}.

\section{Results}\label{sec:results}
\Cref{subsec:bulk} addresses the correlation effects for the  
bulk setup for a semicircular density of states (DOS),
which is realized by the Bethe lattice with infinite coordination number.
The results presented here address the mechanism of gap 
closing as a function of temperature for the bulk HMF\@.
Next, in \cref{subsec:hetero}, we investigate the spectral function of
bilayers made of a metal, a band-insulator, or a Mott insulator
attached to the HMF\@. In particular,
we study the changes induced by stacking a larger number of layers.
Contrary to the Bethe DOS in the bulk case,
for the bi- and multilayers we use the DOS of a square lattice within the layers.

\subsection{Finite temperature behavior of minority spin gap in bulk}\label{subsec:bulk}
In the one-band model \cref{eq:Hubbard-Hamiltonian} a simple way to generate the half-metallic ferromagnetic
state is to introduce a sufficiently strong spin splitting such that 
one spin subband is empty (or full) in the Hartree-Fock (Stoner) picture.
\Cref{subsec:bulk} discusses the results for a homogeneous Hubbard Hamiltonian \cref{eq:Hubbard-Hamiltonian}
of a Bethe lattice with infinite coordination number with half-bandwidth \(D = \SI{1}{eV}\),
spin splitting \(h = \SI{0.5}{eV}\),
on-site energy \(\epsilon - \mu = \SI{1.5}{eV}\),
and on-site interaction \(U = \SI{2}{eV}\).
Difficulties in solving the Hubbard model for such a saturated ferromagnet
are well known\cite{Edwards_1973}.
For the real-frequency results at zero temperature,
the hybridization of the Bethe lattice \(\Delta_\sigma(E) = {(D/2)}^2 G_\sigma(E)\) was discretized using \num{251} bath sites per spin.
We find the ground-state \(\ket{GS}\) (\(T=0\)) to be almost fully polarized (\({n_\downarrow} \sim 10^{-6}, n_{\uparrow} = 0.342\)).
The interacting Green's function and in turn the spectral function were calculated\cite{ganahl2015efficient}
from the time-evolved \(\hat{c}^{(\dagger)}_{\downarrow} \ket{GS}\) and \(\hat{c}^{(\dagger)}_{\uparrow} \ket{GS}\).
This was done using time steps of \SI{0.05}{eV^{-1}},
up to a maximal time of \(t_\text{max}= \SI{150}{eV^{-1}}\).
A linear prediction\cite{Barthel_Schollwock_White_2009,ganahl2015efficient} was performed
for the Green's functions during the last 20 DMFT iterations up to a maximal time of \(t_\text{max} = \SI{1500}{eV^{-1}}\),
so that no dampening of the time series was required.
For the singular-value decompositions, a truncated weight of $t_w = 10^{-10}$
together with a maximal matrix dimension of \num{700} was chosen.
This maximal dimension was reached during the time-evolution of \(\hat{c}^{(\dagger)}_\downarrow \ket{GS}\) at \(t= \SI{100}{eV^{-1}}\).
The truncated weight always remained below \(10^{-8}\).
For the QMC results at finite temperature
we compute the self-energy via the ratio of the two-particle Green's function \(F_\sigma\)
and the one-particle Green's function \(G_\sigma\):\cite{bu.he.98}
\begin{align}
    F_\sigma(\tau - \tau^\prime)\label{eq:two-particle_F}
    &= \expval{\gc_\sigma(\tau)\gcc_{-\sigma}(\tau)\gc_{-\sigma}(\tau)\gcc_\sigma(\tau^\prime)}_{\Seff}
    \\
    \Sigma_\sigma(\iw)
    &= U F_\sigma(\iw)/G_\sigma(\iw).\label{eq:self}
\end{align}
The brackets \(\expval{\cdot}_{\Seff}\) denote the average in the effective impurity model.
This provides more accurate results than the Dyson equation,
such that the Pad\'e analytic continuation\cite{baker_essentials_1975,vi.se.77}
of the self-energy is reasonably accurate.
We calculate the spectral function from the analytically continued self-energy:
\begin{equation}\label{eq:spectral-function}
  A_{\sigma}(E)
  = -\frac{1}{\pi}\Im \int\limits_{-D}^{D}\! \mathrm{d}E^{\prime}
  \frac{\rho(E^{\prime})}{E - E^{\prime} + \tilde{\epsilon}_{\sigma} - \widetilde{\Sigma}_{\sigma}(E)},
\end{equation}
where \(\rho(E)\) is the one-particle density of states of the non-interacting lattice,
and \(\widetilde{\Sigma}(E)\) is the Pad\'e analytic continuation of the self-energy, \cref{eq:self}.

\begin{figure}[htb!]
  \begin{captivy}{\includegraphics[width=\linewidth]{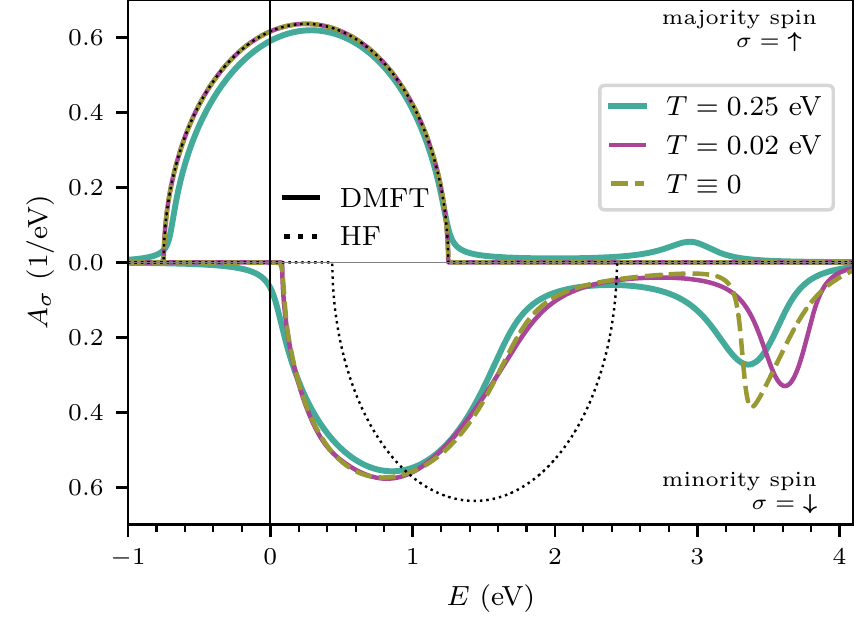}}
    \oversubcaption{0.19, 0.94}{}{subfig:bulk_full}
  \end{captivy}
  \begin{captivy}{\includegraphics[width=\linewidth]{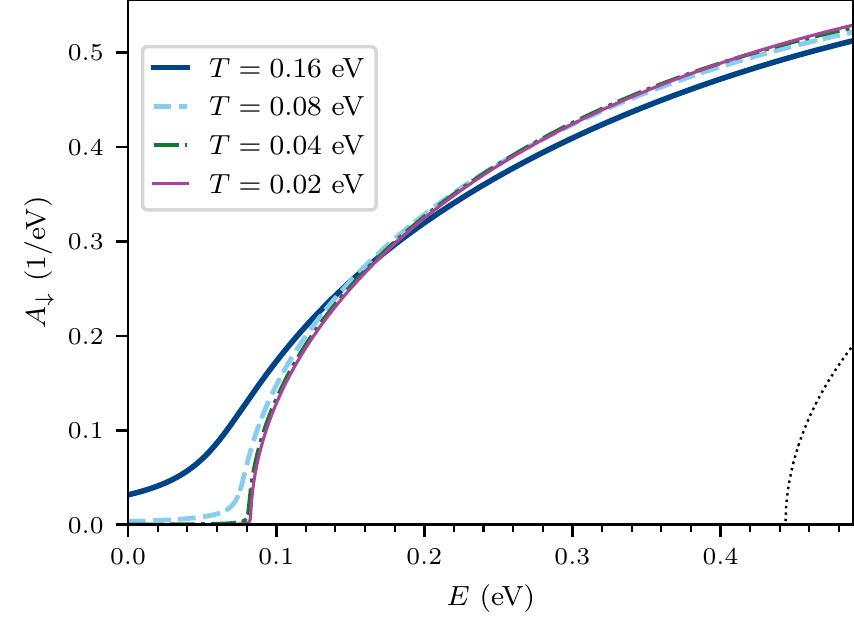}}
    \oversubcaption{0.19, 0.94}{}{fig:nqp_bulk_temperature_dependence}
    \node[anchor=south west,inner sep=0] at (0.48, 0.2) {\includegraphics[scale=1.0]{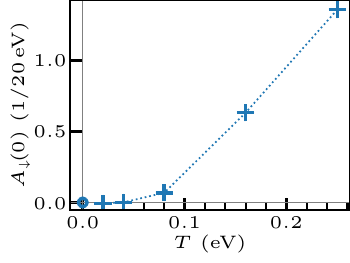}};
    \oversubcaption{0.61, 0.55}{}{subfig:A0_temperature-dependence}
  \end{captivy}
  \vspace*{-0.5cm}
  \caption{%
    \protect\subref{subfig:bulk_full} Spin-resolved spectral function \(A_{\sigma}(E)\) for the bulk half-metal.
    Black dotted lines correspond to the Hartree-Fock (HF) approximation.
    At high temperatures (\(T=\SI{0.25}{eV}\)) the tail of the spectral function \(A_\downarrow(E)\) crosses the Fermi energy,
    while for low enough temperatures (\(T \le \SI{0.02}{eV}\)) the half-metallic gap is preserved.
    \protect\subref{fig:nqp_bulk_temperature_dependence} Evolution of the tail in the minority spin spectral function \(A_{\downarrow}(E)\) with temperature,
    the lowest temperature corresponding to the low \(T\) result of \protect\subref{subfig:bulk_full}.
    Again, the black dotted line shows the HF result.
    The inset \protect\subref{subfig:A0_temperature-dependence} displays the \(T\) dependence
    of the spectral weight \(A_{\downarrow}(E=0)\).
    Crosses indicate finite temperatures shown in \protect\subref{fig:nqp_bulk_temperature_dependence}
    as well as \(T=\SI{0.25}{eV}\);
    the circle corresponds to \(T\equiv 0\).
  }\label{fig:bulk}
\end{figure}
\begin{figure}[htb!]
  \includegraphics[width=\linewidth]{{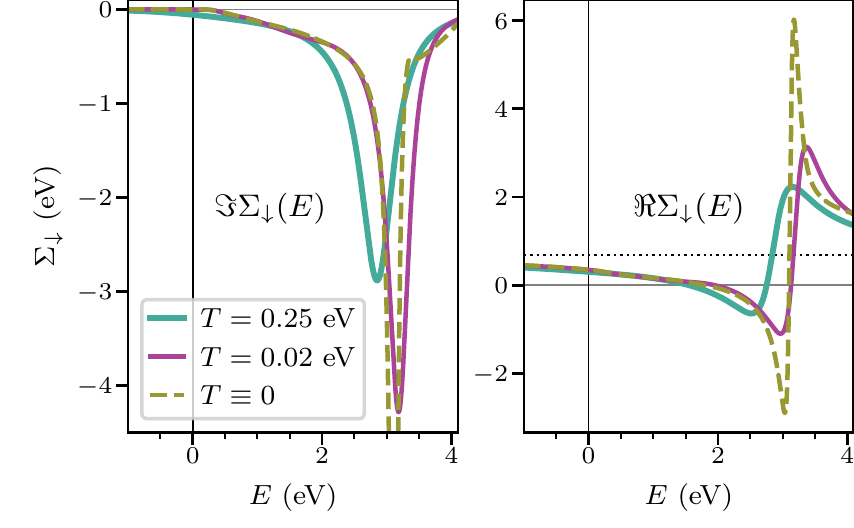}}
  \vspace*{-0.5cm}
  \caption{%
    Imaginary (left) and real (right) part of the down-spin self-energy \(\Sigma_{\downarrow} (E)\) for the bulk half-metal.
    The dotted line for the real part indicates the HF result.
    The peak in \(\Im \Sigma_{\downarrow}(E)\) for \(T \equiv 0\) (truncated in the figure) extends downwards till \SI{-9.1}{eV}.
  }\label{fig:bulk_self-energy}
\end{figure}

\Cref{subfig:bulk_full} displays the results of the DMFT calculations
for zero, low (\(T=\SI{0.02}{eV}\)), and high (\(T=\SI{0.25}{eV}\)) temperature.
The dotted line shows the Hartree-Fock (HF) solution as a reference.
We first discuss the \(T\equiv 0\) spectrum.
As the \(\downarrow\)-spin is completely depleted,
the result for the \(\uparrow\)-spin are nearly identical to the HF result.
The \(\uparrow\)-spin electrons are almost uncorrelated,
the magnitude of the self-energy \(\Sigma_\uparrow(E)\) is negligibly small.
For the \(\downarrow\)-spin we see two main effects of correlations.
First, the size of the gap is reduced compared to the HF approximation.
For low energies, there is a dynamical reduction of the (static) Hartree self-energy (compare \cref{fig:bulk_self-energy}).
Additionally, a many-body satellite appears at \(E \approx \SI{3.5}{eV}\) in \(A_\downarrow(E)\) as shown in \cref{subfig:bulk_full}.
At low temperature \(T=\SI{0.02}{eV}\) the QMC result for the spectral function \cref{eq:spectral-function}
is in good agreement with the real-frequency results for zero temperature.
There is a deviation for the satellite,
however analytic continuation is not expected to resolve features this high in energy well.
At high temperature, \(T=\SI{0.25}{eV}\),
we obtain a tail crossing the Fermi level \(E=0\) shown in \cref{subfig:bulk_full}
which depolarizes the HMF\@.
Due to the tail the \(\downarrow\)-spin is now partially filled, resulting in correlation effects also in the \(\uparrow\)-spin.
The many-body satellite is visible in both spin channels for the high temperature.
Previous calculations\cite{ch.ka.03} used a simplified quantum Monte-Carlo scheme within the so-called exact enumeration technique\cite{georges_dynamical_1996},
therefore results for high temperature (\(T=\SI{0.25}{eV}\)) only were accessible.
Our high $T$ results differ from the previous ones~\cite{ch.ka.03} which
show additional peaks in the spectral function.
In contrast to the previous calculations~\cite{ch.ka.03},
we determine the spectra from the analytically continued self-energy using~\cref{eq:spectral-function}.
In fact, we demonstrate in \Cref{sec:pade} that a Pad\'e analytic continuation of the 
Green's function---instead of the self-energy \cref{eq:self}---causes the appearance of 
these spurious features in the spectral function.

\Cref{fig:nqp_bulk_temperature_dependence} shows the temperature dependence of the 
spectral function for the minority spin, \(A_{\downarrow}(E)\),
in particular its tail crossing the Fermi level.
The highest temperature is \(T=\SI{0.16}{eV}\),
subsequent lines correspond to always half the previous value.
The disappearance of the spectral weight at the Fermi level with decreasing temperature
is apparent.
A specific many-body feature in HMFs is attributed to 
spin-polaron processes~\cite{ir.ka.83}:
the down-spin electron excitations forbidden in the
one-electron description of HMFs arise due to the
superposition of up-spin electron excitations and virtual magnons.
In model calculations the existence of this feature has been shown by 
perturbation-theory arguments for the broad-band 
case~\cite{Edwards_1973} (cf.\ next paragraph), and in the opposite, infinite-$U$
limit~\cite{katsnelson_halfmetallic_2008,ir.ka.83}.
An analytic approximation allows to explore the 
shape of the temperature dependence of the 
spectral function for the minority spins considering
a contact electron-magnon interaction described by
the exchange parameter~\cite{ir.ka.83,ir.ka.85,katsnelson_halfmetallic_2008}.
According to this theory, a 
non-linear temperature dependence is obtained from the competing 
effects of the magnon contribution to the residue 
of the Green's function, $\sim T^{3/2}$, with the shift 
of the band edge states being proportional to $T^{5/2}$.
By a direct fit \(A_{\downarrow}(E=0) \propto T^\alpha\) to
the data in the inset \cref{subfig:A0_temperature-dependence} an
exponent $\alpha$ in the range of $3/2$ to $2$ is obtained.

Considering the perturbation-theory arguments in more detail, we first note that
for a completely depleted down-spin channel as depicted in \cref{subfig:bulk_full} for \(T\equiv 0\) and \(T = \SI{0.02}{eV}\),
it is evident that an added up-electron (or hole) is not subject to interactions.
Therefore, the \textit{up-spin} self-energy \(\Sigma_{\uparrow}(E)\) vanishes.
On the other hand, there is a significant contribution to the \textit{down-spin} self-energy, \(\Sigma_{\downarrow}(E)\),
due to scattering at up-spin electron-hole pairs that arise because of electronic correlations\cite{Edwards_1973,ir.ka.83,ir.ka.85,katsnelson_halfmetallic_2008,oh.ma.16} (while down-spin electron-hole pairs are not possible as the minority spin channel is depleted).
The ferromagnetic instability is triggered by the scattering of the down electron and the up hole,
hence this electron-hole triplet ``bound-state'' can be considered a
magnon~\cite{Edwards_1973,katsnelson_halfmetallic_2008}. 
In perturbation theory, the following expression for the 
imaginary part of the self-energy is found:
\begin{multline}
  \Im \Sigma_{k,\downarrow}(E)
  =-\frac{\pi U^2 n_\uparrow}{N} \sum_{q}(1-f_{\epsilon_{k+q,\uparrow}}+n_{\omega_{q}})
  \\
  \times \delta(E-\epsilon_{k+q,\uparrow}-\hbar \omega_{q})
\end{multline}
where \(\omega_{q} \propto q^2\) is the magnon dispersion;
$f_\epsilon$ denotes the Fermi-Dirac and $n_\omega$ the Bose-Einstein distribution.
As a consequence of the local approximation of DMFT,
the momentum dispersion of the magnons is lost; nevertheless, there is a pole in the magnetic susceptibility corresponding to a local spin flip.
We thus conclude that the DMFT solver includes the scattering of electrons at virtual ``magnons'' (of purely electronic origin), which can be described by diagrams constructed from the local Green's function, and that
the (numerical) local self-energy describes the same type of effective low energy physics as discussed earlier~\cite{Edwards_1973,ir.ka.83,ir.ka.85,katsnelson_halfmetallic_2008}.  
\Cref{fig:bulk_self-energy} presents the self-energy for down-spin electrons
corresponding to the spectral functions shown in \cref{subfig:bulk_full}.
At zero (\(T\equiv 0\)) and low (\(T=\SI{0.02}{eV}\)) temperature the imaginary part of the self-energy \(\Im \Sigma_{\downarrow}\) vanishes at the Fermi level (\(E=0\));
for high temperature (\(T=\SI{0.25}{eV}\)) there is a finite tail, \(-\Im \Sigma_{\downarrow} > 0\), crossing the Fermi level.
The minimum of \(\Im \Sigma(E)\) is located in the energy range \SIrange{3}{3.5}{eV} for the temperatures considered,
slightly below the energies where the satellite in the spectral function is visible in \cref{subfig:bulk_full}.
The satellite is located in the range \(E - \Re \Sigma_{\downarrow}(E) \in (\tilde{\epsilon}_{\downarrow} - D, \tilde{\epsilon}_{\downarrow} +D)\); this range 
is reduced further due to the peak in the imaginary part of the self-energy. As a consequence, the satellites in the spectral functions are found at energies slightly above the peak of the imaginary part of the self-energy.

Furthermore, we investigate the \emph{local} spin-flip susceptibility
which we calculate from the effective impurity model:
\begin{equation}\label{eq:chipm}
\begin{aligned}
  \chi^{+-} (\tau - \tau^{\prime})
  &= \expval{S^{+}(\tau)S^{-}(\tau^{\prime})}_{\Seff}
  \\
  &= \expval{\gcc_{\uparrow}(\tau)\gc_{\downarrow}(\tau)\gcc_{\downarrow}(\tau^{\prime})\gc_{\uparrow}(\tau^{\prime})}_{\Seff},
\end{aligned}
\end{equation}
where \(\Seff\) is the same effective impurity model action from DMFT as in \cref{eq:two-particle_F}.
At zero temperature, the spin-flip susceptibility was obtained directly on the real axis by time-evolving
the matrix-product state \(\ket{\psi} = \ocd_\downarrow \oc_\uparrow\ket{GS}\) using TEBD
and then calculating the overlap \(\chi^{+-}(\tau-\tau') = \braket{\psi(\tau)}{\psi(\tau')}\).
Finite temperature results were sampled with worm-sampling in CT-HYB\cite{wallerberger_w2dynamics_2019};
the analytic continuation to real frequencies was performed using a sparse modeling approach\cite{otsuki_sparse_2017,yo.ot.19}.
\begin{figure}[htb]
  \includegraphics[width=\linewidth]{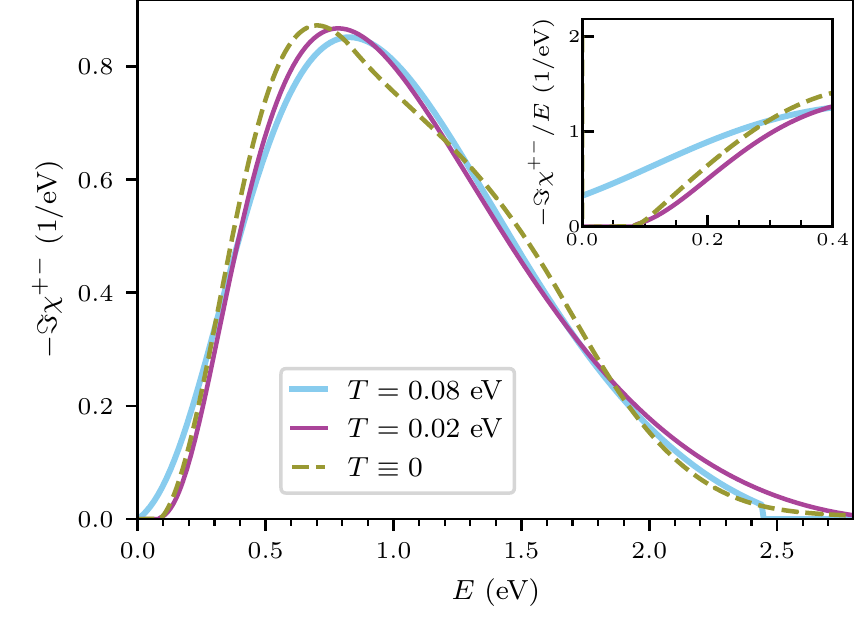}
  \vspace*{-0.5cm}
  \caption{%
    Imaginary part of the local spin-flip susceptibility at zero temperature 
    (dashed green line), and for a selection of finite temperatures (\(T= 0.02\) and \(\SI{0.08}{eV}\): 
    purple and blue solid lines).
    The inset shows the susceptibility divided by energy: \(-\Im \chi^{+-}(E)/E\).
  }\label{fig:spinflip}
\end{figure}
\Cref{fig:spinflip} shows the imaginary part of the susceptibilities \(\chi^{+-}(E)\) 
for different temperatures.
For low and zero temperature, the imaginary part is gapped, i.e.,
it vanishes for a finite region around \(E=0\), in correspondence with
the gapped spectral function shown in \cref{fig:nqp_bulk_temperature_dependence}.
For high-temperatures, on the other hand, we obtain a power-law behavior,
\(\lim_{E\rightarrow0}(-\Im\chi^{+-}(E)/E) > 0\), as visible in the inset of \cref{fig:spinflip};
this is in agreement with the closing of the gap in \cref{fig:nqp_bulk_temperature_dependence}.
All curves have one peak;
the peak position (in energy) seems to slightly increase with temperature.
The real-frequency results show an additional shoulder around \(E \approx \SI{1.5}{eV}\).
In addition, a small satellite is found near \(E \approx \SI{3.5}{eV}\),
outside the area shown.

\subsection{Bi- and multilayers}\label{subsec:hetero}
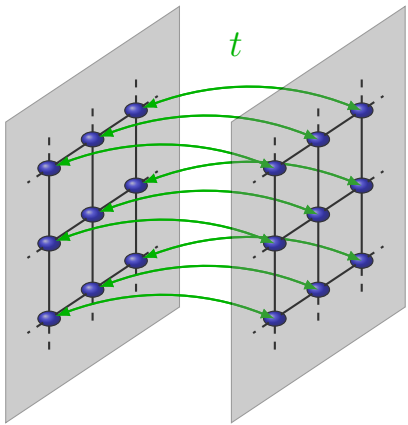
\begin{figure}[h]
  \input{tikz/layer_structure.tikz}
  \caption{Illustration of coupled monolayers of square lattices.
    The in- and inter-layer hopping integrals are indicated.
  }\label{fig:bilayer}
\end{figure}

The starting point is the formulation of the Hamiltonian for the coupled layers. For a given
number of layers \(l\) it has the form:
\begin{equation}\label{eq:ham_bilay}
\begin{aligned}
  \hat{H}
  &= \sum_l \hat{H}_l + \sum_{\langle l, l^\prime\rangle}\hat{H}_{ll'}
  \\
  \hat{H}_l
   &= \sum_{\alpha\beta, \sigma} [{t}^l_{\alpha\beta}+ \tilde{\epsilon}_{l \sigma}\delta_{\alpha\beta}] \ocd_{l\alpha\sigma} \oc_{l\beta\sigma}
  + U_l \sum_{\alpha}  \hat{n}_{l\alpha\uparrow} \hat{n}_{l\alpha\downarrow}
  \\
  \hat{H}_{ll'}
  &= t_{ll^\prime} \sum_{\alpha\sigma} \ocd_{l\alpha\sigma}\oc_{l^\prime \alpha\sigma}
  + \frac{1}{2}\sum_{ll^\prime} \tilde{V}_{ll^\prime} \hat{n}_l \hat{n}_{l^\prime}.
\end{aligned}
\end{equation}
The indices \(\alpha, \beta\) denote sites within a given layer \(l\).
The first term in this Hamiltonian, containing \(\hat{H}_l\), describes isolated layers;
analogous to \cref{eq:Hubbard-Hamiltonian}, this is a sum of single-band Hubbard Hamiltonians.
The second term, a double sum over nearest-neighbor layers (\(\hat{H}_{ll^\prime}\)),
contains the hopping between adjacent layers as well as the inter-layer Coulomb interaction.
The latter
is treated, for simplicity, within a mean-field approximation:
\begin{equation}
\frac{1}{2}\sum_{ll^\prime} \widetilde{V}_{ll^\prime} \hat{n}_l \hat{n}_{l^\prime}
  \approx 
  \sum_l V_l \hat{n}_l.
\end{equation}
This is equivalent to using Poisson's equation\cite{chen_electronic_2007,hale_manybody_2012}
to determine the potential self-consistently.
Within the layers, we consider a two-dimensional square lattices as
depicted in \cref{fig:bilayer}.
The density of states in a single 
layers has a half-bandwidth of \(D=\SI{1}{eV}\), which corresponds to an 
in-plane hopping \(t^{l}_{\alpha \beta} =\SI{0.25}{eV}\) for nearest-neighbors \(\alpha, \beta\).
The inter-layer hoppings are chosen as \(t_{l, l+1} = \SI{0.5}{eV}\).
For the remainder of \cref{subsec:hetero}, we fix the temperature at \(T = \SI{0.16}{eV}\).

\subsubsection{Bilayers}\label{sec:bilayer}
The systems studied next consist of two coupled layers;
one of the layers (\(l=1\)) is half-metallic and the other (\(l=2\)) is either a metal, 
a band insulator, or a Mott insulator.
The half-metallic layers have the same parameters as in \cref{subsec:bulk}:
\(h_l = \SI{0.5}{eV}\), \(\epsilon_{l} = \SI{-1.5}{eV}\), and \(U_l = \SI{2}{eV}\).
We fix the filling of the bilayer to match the sum of the fillings of the isolated layers \(n^\text{iso}_l\);
the HMF layer contributes a filling of \(n^{\text{iso}}_1 = 0.355\).
The nearest-neighbor inter-layer hopping \(t_{12}= t_{21} =t\) couples the layers.

{}

In the absence of interactions, $U_{l}=0$ and $V_l=0$,
and in the presence of a splitting field \(h_1\) acting only on the HMF layer \(l=1\),
the energy spectrum shows bonding \((E^{-}_\sigma(k_\parallel))\)
and anti-bonding \((E^{+}_\sigma(k_\parallel))\) sub-bands:
\begin{equation}
\begin{aligned}
  E^\pm_\sigma(k_\parallel)
  &= \epsilon(k_{\parallel}) + \bar{\epsilon}_\sigma \pm \sqrt{\bar{\epsilon}_\sigma^2+\Delta^2_\sigma}
  \eqqcolon \epsilon(k_{\parallel}) + \epsilon_\sigma^\pm
  \\
  \epsilon(k_{\parallel}) &= -2t (\cos k_x + \cos k_y)
  \\
  \bar{\epsilon}_\sigma &= (\tilde{\epsilon}_{1 \sigma} + \tilde{\epsilon}_2)/2
  \\
  \Delta_\sigma^2 &= t^2 - \tilde{\epsilon}_{1 \sigma} \tilde{\epsilon}_2
  \\
  \tilde{\epsilon}_{1 \sigma} &= \tilde{\epsilon}_1 + \sigma h_1.
\end{aligned}
\end{equation}
For the Green's functions of the layers \(l = 1,2\) we get
\begin{multline}
    G^0_{ll\sigma}(z, \epsilon(k_\parallel))
    \\
    =\frac{1}{\epsilon^+_\sigma - \epsilon^-_\sigma}
    \left[\frac{\tilde{\epsilon}_{l\sigma}- \epsilon^-_\sigma}{(z - \epsilon(k_\parallel) - \epsilon^+_\sigma)}
    \right.
    -\left. \frac{\tilde{\epsilon}_{l\sigma} - \epsilon^+_\sigma}{(z - \epsilon(k_\parallel) - \epsilon^-_\sigma)}
    \right].
\end{multline}
The magnetic field ($h_1$) splits the two spin channels. 

\begin{figure}[htb!]
  \begin{captivy}{\includegraphics[width=\linewidth]{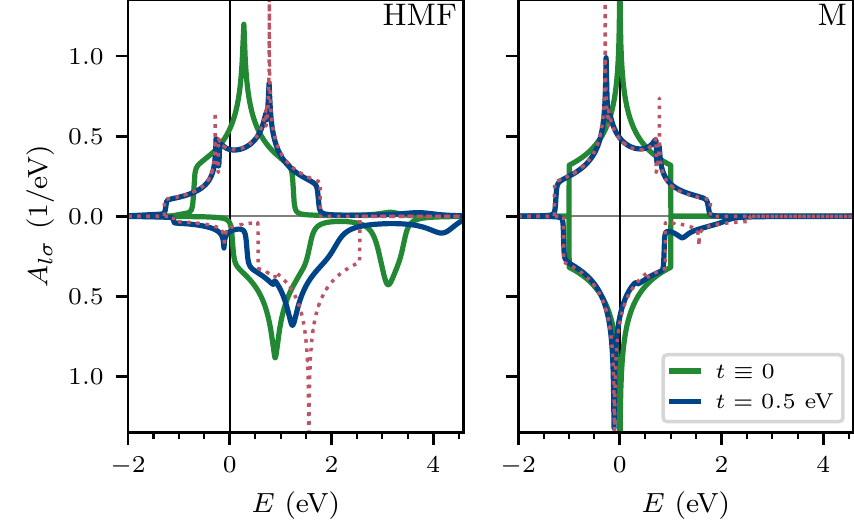}}
    \oversubcaption{0.19, 0.92}{}{subfig:A_sq_hmf1_m1:hmf}
    \oversubcaption{0.64, 0.92}{}{subfig:A_sq_hmf1_m1:m}
  \end{captivy}
  \vspace*{-0.5cm}
  \caption{%
    Spin-resolved spectral function \(A_{l \sigma}(E)\) for one HMF layer \protect\subref{subfig:A_sq_hmf1_m1:hmf}
    interfaced with one metallic (M) layer \protect\subref{subfig:A_sq_hmf1_m1:m}.
    The solid lines are DMFT (CT-HYB), and the dotted lines the HF results (\(t=\SI{0.5}{eV}\)).
    The green lines show the spectral function for isolated layers (\(t\equiv 0\)).
  }\label{fig:A_sq_hmf1_m1}
\end{figure}
\Cref{fig:A_sq_hmf1_m1} shows the spectral functions of the bilayer heterostructure with one HMF layer coupled to a metallic layer~(M).
The metallic layer \(l=2\) is non-interacting, \(U_{2} = 0\),
non-magnetic, \(h_{2} = 0\),
and half-filled, \(\epsilon_{2} = 0, n_{2}^{\text{iso}} = 1\);
these values imply a chemical potential of \(\mu = \SI{-0.078}{eV}\).
Both layer spectral functions \(A_{l \sigma}(E)\), \(l =1,2\) are metallic;
the gap in the minority channel of the HMF layer \(l=1\) closes.
The essential physics is the charge transfer between the half-metallic 
and the metallic layer, which increases the filling
in the minority spin channel of the half-metal that closes the gap.
This effect also occurs in the absence of interactions. 

\begin{figure}[htb!]
  \begin{captivy}{\includegraphics[width=\linewidth]{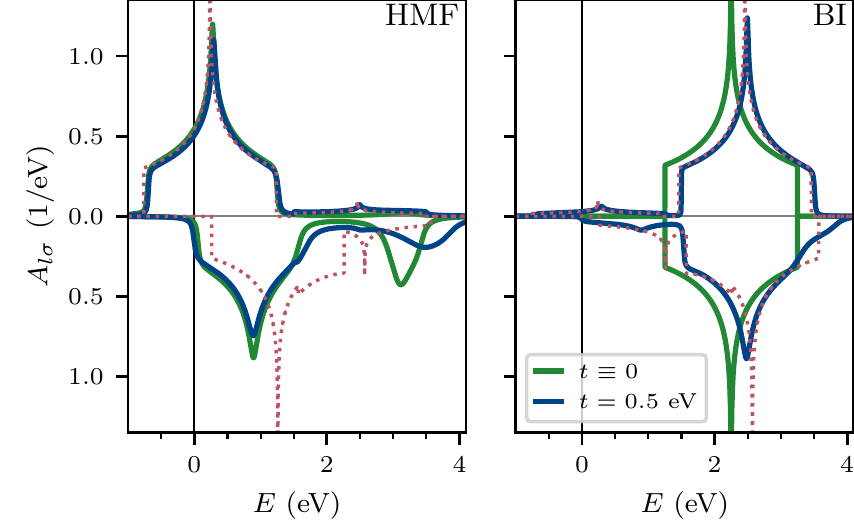}}
    \oversubcaption{0.19, 0.92}{}{subfig:A_sq_hmf1_bi1:hmf}
    \oversubcaption{0.64, 0.92}{}{subfig:A_sq_hmf1_bi1:bi}
  \end{captivy}
  \vspace*{-0.5cm}
  \caption{%
    Spin-resolved spectral function \(A_{l\sigma}(E)\) for one HMF layer \protect\subref{subfig:A_sq_hmf1_bi1:hmf}
    interfaced with one band insulating (BI) layer \protect\subref{subfig:A_sq_hmf1_bi1:bi}.
    The solid lines are the DMFT (CT-HYB), and the dotted lines the HF results (\(t=\SI{0.5}{eV}\)).
    The green lines show the spectral function for isolated layers (\(t\equiv 0\)).
  }\label{fig:A_sq_hmf1_bi1}
\end{figure}
\Cref{fig:A_sq_hmf1_bi1} shows the spectral function of a bilayer structure of a HMF
layer interfaced with a band-insulating (BI) layer.
The band-insulating layer \(l=2\) is non-interacting, \(U_2=0\),
non-magnetic, \(h_2 = 0\),
and completely empty, \(\epsilon_{2} = \SI{-2.25}{eV}\), \(n^{\text{iso}}_2 = \num{5e-5}\);
these values imply a chemical potential of \(\mu = \SI{-0.129}{eV}\).
The layer-resolved spectral functions
show that the disappearance of the minority spin half-metallic gap is
due to the interactions in the half-metallic layer.
According to the HF solution of the bilayer,
both layers show a gap for down-spin electrons,
cf.\ the dotted lines in \cref{subfig:A_sq_hmf1_bi1:hmf,subfig:A_sq_hmf1_bi1:bi}.
The proximity to the correlated
HMF layer causes the appearance of electronic states around the Fermi level of the 
band insulator. 
The many-body induced tail in the HMF is enhanced, decreasing the polarization 
of the HMF layer further.

\begin{figure}[htb]
  \begin{captivy}{\includegraphics[width=\linewidth]{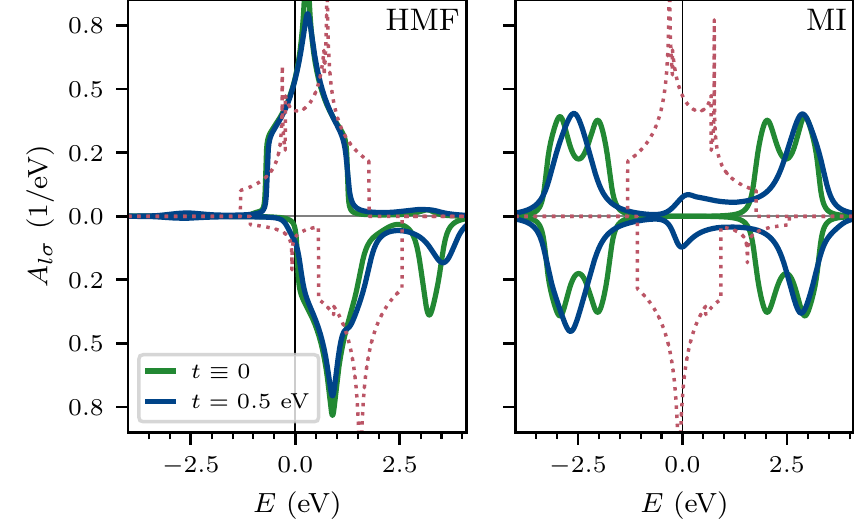}}
    \oversubcaption{0.19, 0.92}{}{subfig:A_sq_hmf1_mi1:hmf}
    \oversubcaption{0.64, 0.92}{}{subfig:A_sq_hmf1_mi1:mi}
  \end{captivy}
  \begin{captivy}{\includegraphics[width=\linewidth]{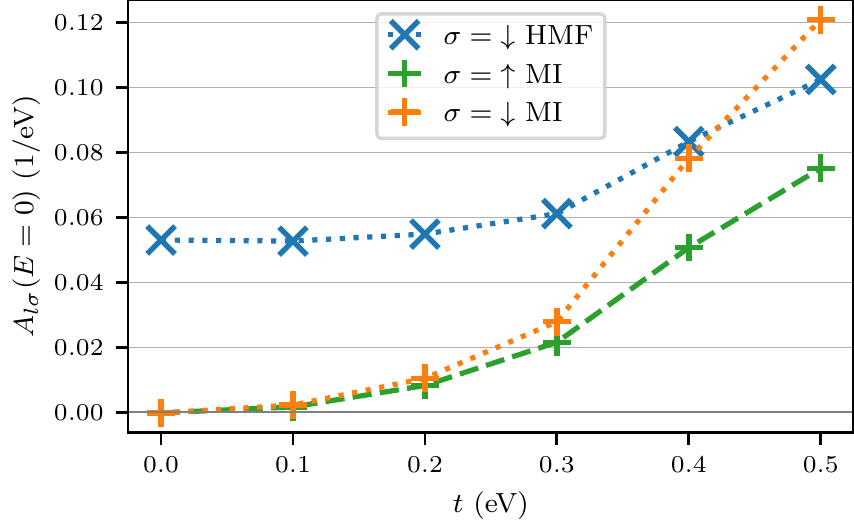}}
      \oversubcaption{0.19, 0.89}{}{fig:A_sq_hmf1_mi1_tX}
  \end{captivy}
  \vspace*{-0.5cm}
  \caption{%
    Spin-resolved spectral function \(A_{l\sigma}(E)\) for one HMF layer \protect\subref{subfig:A_sq_hmf1_mi1:hmf}
    interfaced with one Mott insulator (MI) layer \protect\subref{subfig:A_sq_hmf1_mi1:mi}.
    The solid lines are the DMFT (CT-HYB), and the dotted lines the HF results (\(t=\SI{0.5}{eV}\)).
    The green lines show the spectral functions for isolated layers (\(t\equiv 0\)).
    \protect\subref{fig:A_sq_hmf1_mi1_tX}
    Spectral weight at the Fermi level \(A_{l\sigma}(E=0)\) as function
    of the hopping \(t\) between the layers.
  }\label{fig:A_sq_hmf1_mi1}
\end{figure}

\Cref{fig:A_sq_hmf1_mi1} shows the spectral functions of the bilayer formed
by interfacing the HMF layer and a Mott insulating (MI) layer.
Electrons in the MI layer are
subject to a considerable interaction, \(U_2 = \SI{5}{eV}\), 
no magnetic splitting, \(h_2 = 0\),
and for the layer occupation the half-filled case 
(\(\epsilon_{2}= 0\), \(n^\text{iso}_2 = 1\)) is considered;
for these parameters, the chemical potential is \(\mu = \SI{0.013}{eV}\).
At the level of HF this corresponds to the interface between the half-metallic 
and the ordinary metallic layer as both spectral functions
show states at and around the Fermi level.
Within the insulating layer, \cref{subfig:A_sq_hmf1_mi1:mi}, the splitting into lower and 
upper Hubbard bands is visible (separated by \(\approx U_{2}\)).
The proximity to the HMF layer induces a slightly spin-polarized quasiparticle peak (QP)
located at the Fermi level of the MI layer. 
In contrast, the isolated Mott layer, \(t_{ll^{\prime}} \equiv 0\), shows no QP peak for these parameters.\cite{fu.he.06,ka.ok.07}  
In order to study the polarization of the QP peak we 
performed calculations increasing the magnitude of \(U_2\) starting from \(U_2 = \SI{1}{eV}\).

\begin{figure}[htb]
  \begin{captivy}{\includegraphics[width=\linewidth]{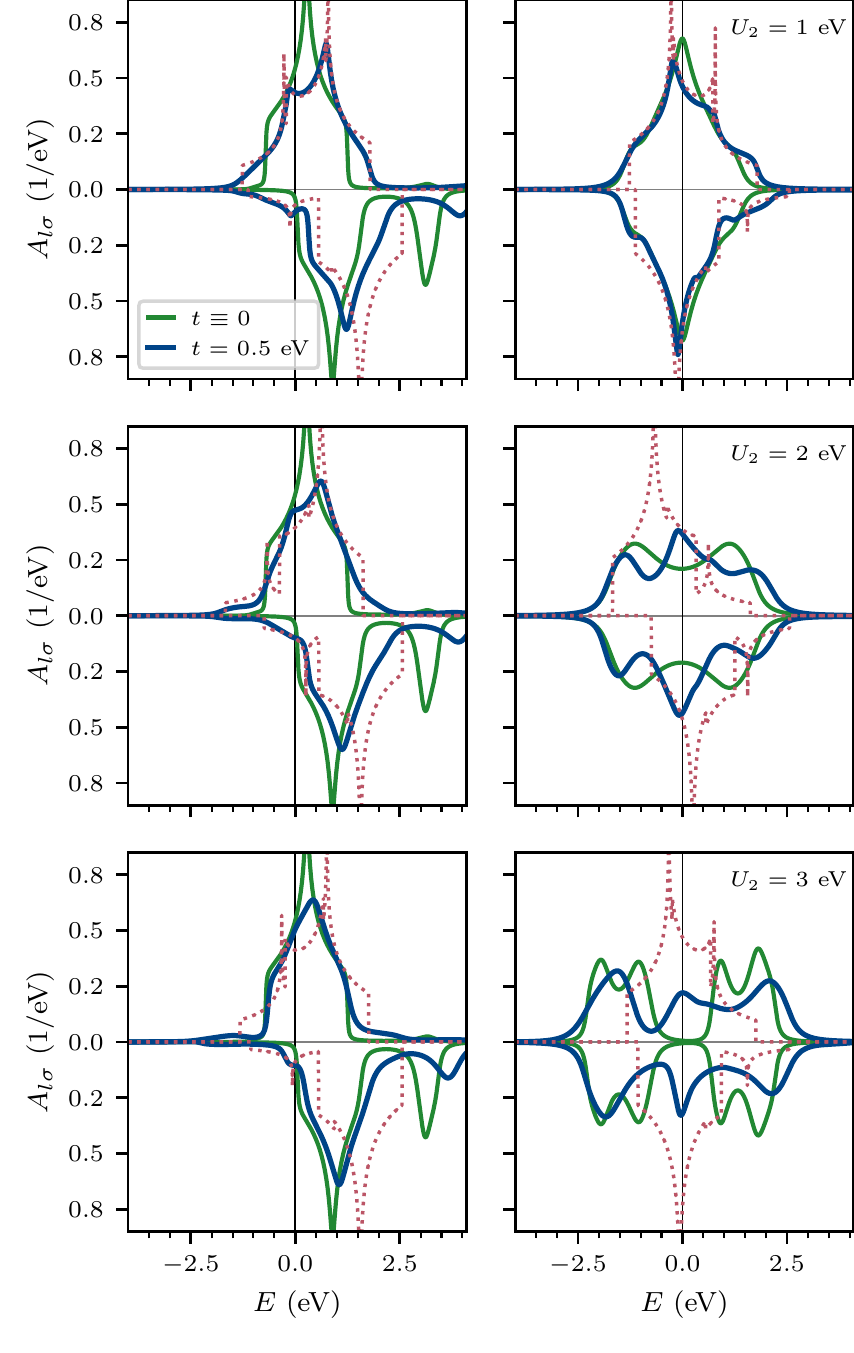}}
    \oversubcaption{0.19, 0.97}{}{subfig:A_sq_hmf1:U1}
    \oversubcaption{0.19, 0.65}{}{subfig:A_sq_hmf1:U2}
    \oversubcaption{0.19, 0.33}{}{subfig:A_sq_hmf1:U3}
  \end{captivy}
  \vspace*{-0.5cm}
  \caption{%
    Spin-resolved spectral function \(A_{l\sigma}(E)\) for one HMF layer interfaced with one layer of different interacting strengths \(U_{2}\).
    The solid lines are the DMFT (CT-HYB), and the dotted lines the HF results (\(t=\SI{0.5}{eV}\)).
    The green lines show the spectral function for isolated layers \(t \equiv 0\).
  }\label{fig:A_sq_hmf1_UX}
\end{figure}

In \cref{fig:A_sq_hmf1_UX} we present the spectral function obtained 
for fixed parameters of the HMF layer (\(U_1 = \SI{2}{eV}, \epsilon_{1} = -\SI{1.5}{eV}, h_1 = \SI{0.5}{eV} \)),
while increasing the strength of the Hubbard parameter \(U_2= 1,2,\SI{3}{eV}\) 
towards a Mott insulator in the adjacent layer, \(l=2\) (\(\epsilon_{2} = 0, h_{2} = 0\)).
The quasiparticle peak and the lower and upper Hubbard bands are already seen for \(U_2=\SI{2}{eV}\) in \cref{subfig:A_sq_hmf1:U2},
their separation increases with increasing \(U_2\). 
The spectral function of the HMF layer shows, besides the expected satellite at 
about \SI{3.5}{eV}, some additional spectral weight corresponding to the position 
of the lower Hubbard band of the Mott insulating layer.
Likewise, at higher energies at the position of the upper Hubbard band a shoulder 
in the spectral function of the HMF layer is visible.
Contrary to the homogeneous single layer, where increasing \(U_{2}\) leads to a sharpening of the QP feature,
the spectral weight induced by the charge-transfer seems to overlay the QP\@.
While the spectral weight around the Fermi level decreases with increasing \(U_2\),
it persists even for values as large as \(U_2 = \SI{10}{eV}\). 
Accordingly, the double occupation of the MI layer is not completely suppressed in the bilayer case:
while increasing the interaction \(U_{2}\) reduces it,
the double occupation is larger than in the isolated MI layer case.

We point out that we do not expect a strict Mott
transition in the sense of a vanishing quasiparticle
weight, respectively of a divergent effective mass.
Instead, the mutual doping of Mott and HMF layer leads
to metallic behavior of the whole bilayer, similarly 
as discussed previously\cite{he.ko.08,no.ma.11}.
Thus the system favors a certain amount of charge 
fluctuations, and the hopping between the layers is never
renormalized to zero.
Such a behavior has been coined ``electronic reconstruction.''\cite{ok.mi.04b}
The common feature of these results indicates that the
transfer of charge between the layers is a general 
phenomenon that produces metallic interfaces.

\subsubsection{Half-metallic and Mott insulating multilayers}\label{sec:multilayers}
In the following, we scale up the system size and consider a heterostructure 
made up of five HMF layers coupled to five Mott insulator layers.
We consider open boundary conditions.
In order to preserve the half-metallic gap, we scale the on-site parameters of the half-metallic layers by a 
factor of two in comparison to the previous bilayer calculations, \cref{sec:bilayer}:
\(U_l = \SI{4}{eV}\), \(h_l = \SI{1}{eV}\), and \(\epsilon_{l} - \mu = \SI{-3}{eV}\) for all
layers \(l \in \{1, \dots 5\}\).
For the Mott insulating layers, we choose the same Hubbard interaction, \(U_l = \SI{5}{eV}\), for all 
remaining layers  \( l \in \{6, \dots 10\}\).

\begin{figure*}
  \begin{captivy}{\includegraphics[width=\linewidth]{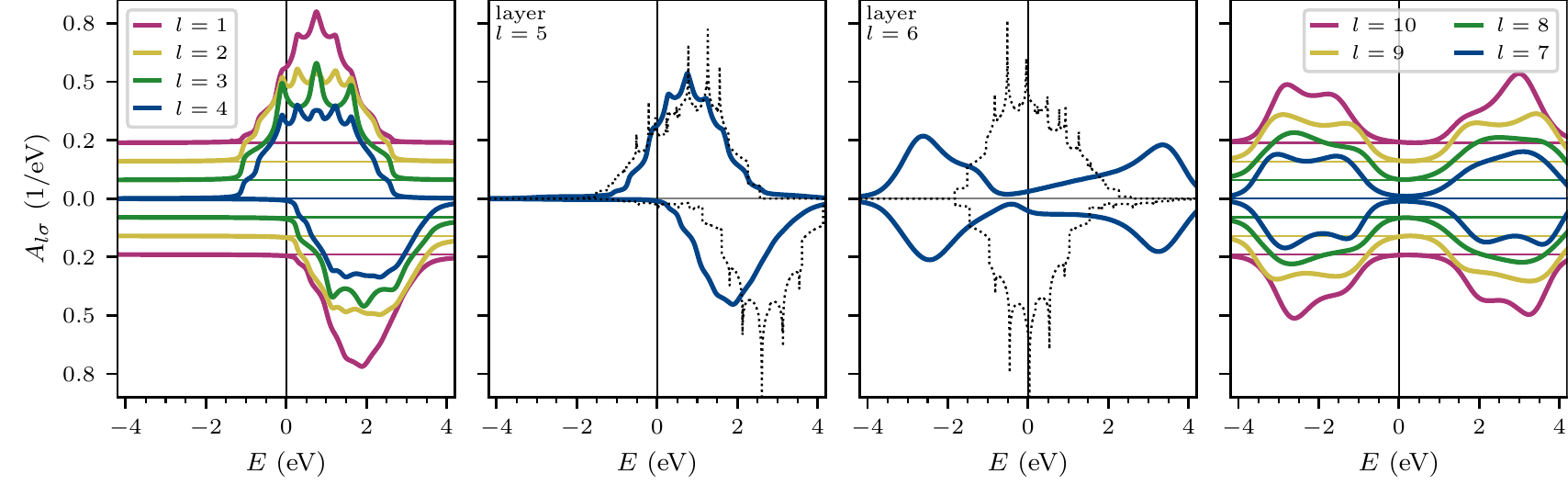}}
    \oversubcaption{0.100, 0.25}{}{subfig:A_sq_hmf5_mi5:HMF}
    \oversubcaption{0.335, 0.25}{}{subfig:A_sq_hmf5_mi5:l5}
    \oversubcaption{0.570, 0.25}{}{subfig:A_sq_hmf5_mi5:l6}
    \oversubcaption{0.805, 0.25}{}{subfig:A_sq_hmf5_mi5:MI}
  \end{captivy}
  \vspace*{-0.5cm}
  \caption{%
    Spin-resolved spectral function, \(A_{l\sigma}(E)\), for  five HMF layers
    interfaced with five MI layers.
    (a) shows the results for the first four layers (HMF);
    the spectra are shifted, their respective baselines are plotted in the same color.
    Panels (b) and (c) represent the HMF interface layer, \(l=5\), and the MI interface layer, \(l=6\),
    respectively; in both cases, the HF results are included as dotted lines for 
    easy reference.
    The results for the last four MI layers are shown in (d); again the spectra are shifted.
    The spectral functions are evaluated by analytic continuation,
    \(E \rightarrow E + \SI{0.04j}{eV}\).
  }\label{fig:A_sq_hmf5_mi5}
\end{figure*}
\Cref{fig:A_sq_hmf5_mi5} shows the layer-resolved spectral function, 
\(A_{l\sigma}(E)\), for this setup.
The many-body effects in the half-metallic layers 
(\cref{subfig:A_sq_hmf5_mi5:HMF,subfig:A_sq_hmf5_mi5:l5}) are qualitatively the same as in bulk:
we observe a dynamical reduction of the Hartree part of self-energy,
a tail crossing the Fermi level,
and a satellite at \(E \approx \SI{7}{eV}\).
Approaching the interface, the satellite shifts to slightly higher energy \(E \approx \SI{7.2}{eV}\).
Within the HF approximation, the layers \(l=6, \dots 10\) (i.e., on the MI side,
\cref{subfig:A_sq_hmf5_mi5:l6,subfig:A_sq_hmf5_mi5:MI})
are found to be metallic, as to be expected; 
in addition, the charge-transfer at the interface introduces small weight at the gap 
in the interface layer on the HMF side, \(l = 5\)
(\cref{subfig:A_sq_hmf5_mi5:l5}).

Within DMFT, we see that the spectral weight in the MI surface layer, \(l=6\),
is strongly suppressed around the Fermi level; however, the layer remains metallic
despite the strong interaction. On the other hand, a significant shift of 
spectral weight towards the Fermi level is apparent in the down-spin channel of the
HMF interface layer, \(l=5\).
Nevertheless, the polarization in this layer (\(l=5\)) 
is close to the polarization obtained within HF\@.
The closing of the Mott gap observed in the interface layer on the MI side is similar
to the bilayer, cf.\ \cref{fig:A_sq_hmf1_mi1}.
This effect has the range of two layers, at \(l=8\) the gap is apparent again.
The short range of penetration is in agreement with the paramagnetic case for a metal-MI interface\cite{he.ko.08}.
The minimum of the spectral function \(A_{6\sigma}\), \cref{subfig:A_sq_hmf5_mi5:l6}, of the MI interface layer shifts from zero energy
to roughly \(E \approx \SI{-1}{eV}\) for both spin channels.
Contrary to the bilayer result, \cref{fig:A_sq_hmf1_mi1}, there is
no QP peak at \(E=0\), neither at the interface nor in the subsequent MI layers.
Surprisingly, the spectral function 
shows a shoulder at the Fermi level for the down-spin only.
Next we include the long-ranged Coulomb repulsion in mean-field approximation
for this multilayer setup.\cite{ok.mi.04,ka.da.06,chen_electronic_2007,hale_manybody_2012,ue.ka.12}
We apply the algorithm described in Ref.~\onlinecite{chen_electronic_2007,hale_manybody_2012};
the formula for the potential reads
\begin{equation}\label{eq:V_l}
  V_{l}(\{n_{l}\})
  = - \sum_{m}(n_{m} - n_{m}^{\text{bulk}})
  \sum_{n=\min(m, l) + 1}^{\max(m, l)} (\eschot_{n} + \eschot_{n-1})
\end{equation}
with the layer occupation \(n_{l} = n_{l\downarrow} + n_{l\uparrow}\);
the material parameter, \(\eschot\), can be related to the screening
length, as discussed previously.\cite{chen_electronic_2007,hale_manybody_2012}
We use, however, a different update scheme to solve the Poisson equation,
which avoids
the thousands of iterations\cite{chen_electronic_2007,hale_manybody_2012}
necessary with a naive mixing scheme.
Instead, after every DMFT iteration we temporarily fix the self-energy, \(\Sigma_{l \sigma}(\iw)\), to a self-consistent potential \(V_{l}\).
We start from the occupation numbers
\(n_{l}\) (e.g., given by the last DMFT iteration, or the non-interacting result).
From the occupations, we calculate the potential using the above
equation, 
\begin{equation}
  \vec{V} = \vec{V}(\vec{n}),
\end{equation}
where we introduce the vector notation \(\vec{V} = \{V_{l}\}\), \(\vec{n} = \{n_{l}\}\).
Given the potential and the self-energy, we can calculate a new Green's function,
\begin{equation}
  \vec{G}(\iw) = \vec{G}(\vec{V},\vec{\Sigma}(\iw), \iw),
\end{equation}
with the vectors \(\vec{G}(\iw) = \{G_{ll \sigma}(\iw)\}\) and \(\vec{\Sigma}(\iw) = \{\Sigma_{l \sigma}(\iw)\}\).
From the Matsubara sum of the Green's function we then calculate new occupations,
giving us the self-consistency equation
\begin{equation}\label{eq:n_self-consistency}
  \vec{n} = \vec{n}[\vec{G}] = \vec{n}[\vec{n}, \vec{\Sigma}].
\end{equation}
After every DMFT step, we solve for the self-consistent charge \(n_{l}\), \cref{eq:n_self-consistency},
and therefore for a self-consistent potential \(V_{l}\) for the given self-energy.
This method significantly reduces the number of required DMFT iteration,
however, it introduces costs for solving \cref{eq:n_self-consistency} after every iteration.
The main cost of \cref{eq:n_self-consistency} is the evaluation of the Green's function matrix.
Numerically, it is more efficient
to solve the equivalent root-search problem, \(\vec{r}(\vec{n}^*) = 0\), for the function 
\begin{equation}\label{eq:n_root}
  \vec{r}(\vec{n}) = \vec{n}[\vec{n}, \vec{\Sigma}] - \vec{n}.
\end{equation}
A Newton-Krylov solver\cite{knollJacobianfreeNewtonKrylov2004}, as implemented in Ref.~\onlinecite{2020SciPy-NMeth},
is found to be most suitable for this problem.
Furthermore, we also include the search for the chemical potential, \(\mu\),
necessary for fixing the total charge and therefore guaranteeing charge neutrality,
\(\sum_{l} n_l = \sum_{l} n^{\text{bulk}}_{l}\), when performing the root search, \cref{eq:n_root}.
This is easily implemented using the modified equation
\begin{equation}\label{eq:n_root_mu}
  \tilde{\vec{r}}(\mat{n}, \mu)
  = \begin{pmatrix}
  \vec{r}(\vec{n})
  \\
  \sum_{l}(n_{l}[\mat{n}, \mat{\Sigma}, \mu] - n^{\text{bulk}}_{l})
  \end{pmatrix}
\end{equation}
where we append a row for the difference in total charge to the vector-valued function \(\vec{r}(\vec{n})\).\footnote{%
    We can formulate an equivalent root-search \(\vec{\tilde{r}}(\vec{V}, \mu)\) problem to \cref{eq:n_root_mu} starting from the potential \(\vec{V}\) instead of the occupation \(\vec{n}\).
}
We fix the material parameter to the following selection of representative values: \(\eschot_{l}=\text{\SIlist{0.2;1.0;5.0;25}{eV}}\).
The bulk occupations are \(n_{l} = 0.205\) for the HMF layers, \(l \in \{1, \dots 5\}\),
and half-filling (\(n_{l} = 1\)) for the MI layers, \(l \in \{6, \dots 10\}\).

\begin{figure}[htbp]
  \includegraphics[width=\linewidth]{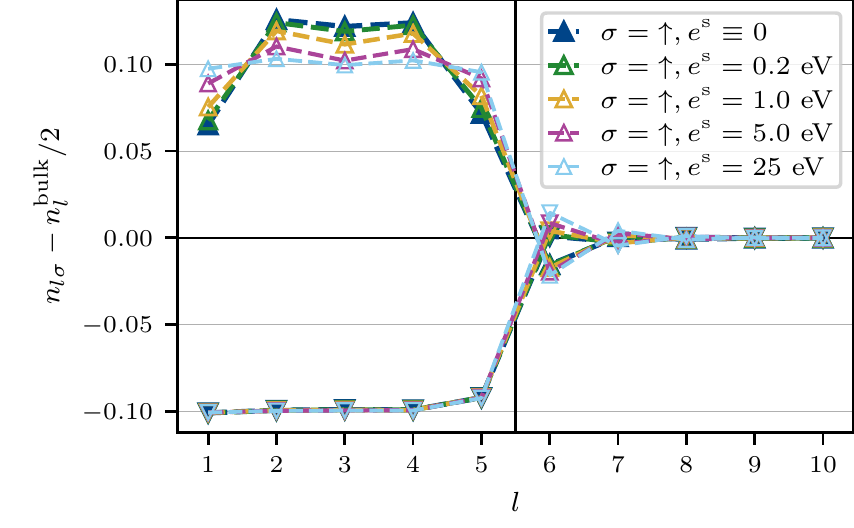}
  \vspace*{-0.5cm}
  \caption{%
    Change in spin-resolved occupations due to inter-layer Coulomb mean-field potential, \(V_l\) \cref{eq:V_l}.
    The graph shows the deviation of the spin-resolved
    occupation of the layers from half the bulk occupation \(n^\text{bulk}_l/2\).
    The blue markers show the occupations for \(\eschot \equiv 0\) and therefore \(V_l \equiv 0\),
    while for the other markers (\(\eschot \neq 0\)) the self-consistent mean-field potential is included.
    The vertical bar indicates the interface between the HMF region (\(l \in \{0, \dots 5\}\))
    and the MI region (\(l \in \{6, \dots 10\}\)).
    The upward triangles represent the up-spin \(\sigma={\uparrow}\),
    the downward triangles the down-spin \(\sigma={\downarrow}\).
  }\label{fig:n_sq_hmf5_mi5_pois}
\end{figure}
\Cref{fig:n_sq_hmf5_mi5_pois} shows the spin-resolved occupation for the multilayer,
without (blue) and including the long-ranged Coulomb repulsion.
The magnitude of \(\eschot\) mainly affects the occupation of the metallic spin-channel \(\sigma={\uparrow}\),
due to charge fluctuations in this channel,
while the rest is nearly invariant with respect to inclusion of long-ranged effects.
Likewise the spectral function is nearly identical to \cref{fig:A_sq_hmf5_mi5}.
We see, however, an increase of the magnitude of the proximity-induced local magnetic moment \(n_{l\uparrow} - n_{l\downarrow}\)
in the Mott layer at the interface (\(l=6\)).

\section{Conclusion}\label{sec:conclusion}

In summary, we have presented detailed model studies 
for the spectral properties of bulk half-metallic ferromagnets (HMFs)
as well as for bi- and multilayers containing half-metallic ferromagnets. Dynamical
mean-field theory has been employed to describe 
the local correlations 
between charge carriers, while
a mean-field approach was used to include the
long-ranged Coulomb interactions. 

Our numerical results show that the correlation-induced
tails in the vicinity of the Fermi level in bulk HMFs
are significantly reduced at zero temperature, in 
agreement with analytical predictions~\cite{katsnelson_halfmetallic_2008}.
On the other hand, for bi- and multilayers we find
an enhancement of the tail contribution at the half-metallic side,
as well as coherent quasiparticle states on the Mott insulating side.
In the multilayers these mobile carriers are confined to a relatively narrow region at the interface.
Furthermore, the Fermi liquid states at the interface
reduce the full spin polarization characteristic for
bulk HMFs. Note that the formation of Fermi liquid 
states at such interfaces is similar to the LAO/STO 
interfaces, which have been theoretically~\cite{br.ti.10,pa.ko.12,pa.ko.12_2,pa.ko.13} studied and experimentally observed~\cite{ohtomo_artificial_2002,ohtomo_highmobility_2004} some time ago; 
however, to the best of our knowledge, such effects have
not been studied for heterostructures containing HMFs. 

On the technical side, we have demonstrated that the
real-space DMFT allows, in a rather transparent way, the
inclusion of long-ranged Coulomb interactions via the
Poisson equation. In this approach, the charge
distribution in the presence of strong short-range
interactions and spatially inhomogeneous hoppings is 
determined self-consistently. In contrast to previous 
implementations of the R-DMFT~\cite{freericks_dynamical_2004}, 
we use the Hubbard model and a state-of-the-art 
CT-QMC~\cite{gull_continuoustime_2011} implementation
for the impurity solver. 
The Poisson equation is solved as an effective
one-dimensional problem in combination with the R-DMFT
self-consistency condition as discussed above. For the 
bi- and the multilayer setup, we have considered the 
case where the layers can be modeled as square lattices. 
For our bilayer setup, we have considered a half-metallic 
monolayer in contact with either a metal, a band or a
Mott insulator. We have seen that charge reconstruction
at the interface causes the existence of metallicity, 
even in the presence of large Hubbard $U$ parameters at 
the Mott insulator layer. In the R-DMFT analysis the
HMF/MI bilayers are Fermi liquids with well defined 
quasiparticles, thus the present approach offers a way 
to access Fermi liquid quantities on the basis of a microscopic model.

On the experimental side, most studies have concentrated
on the question of whether the half-metallic properties extend to the surface or interface. 
Using DFT calculations, a genuine half-metallic 
interface of NiMnSb with InP and CdS 
has been obtained only for the anion terminated (111) direction~\cite{wi.gr.01}.  
Interfaces of semi-Heuslers NiMnSb or NiMnSi with large 
gap insulators such as MgO have been also studied~\cite{zh.ec.14}. A high spin polarization
has been obtained only under the prerequisite of 
structural optimization~\cite{zh.ec.14}. 

However, the microscopic origin of the HMF/Mott
insulator interface has never been addressed. In this
context, we thus considered a minimal model in which 
half-metallic layers are in contact with correlated 
insulator layers. 
We solved the corresponding Hubbard Hamiltonian in
Hartree-Fock (HF) approximation and beyond using
dynamical mean-field theory (DMFT). Within the HF 
approximation, when crossing the interface from the HMF
side into the metallic side, we find a sharp transition,
i.e., the half-metallic layer is followed directly by a
metallic layer.
In contrast, by including dynamical correlations within
DMFT, we find a continuous transition from the 
half-metallic region through a pseudo-gapped interface
into an insulating region.
Our simplified model thus indicates that 
a high spin-polarization within the interface region 
can be preserved 
in the presence of correlated Mott insulators.

\begin{acknowledgments}
The zero-temperature calculations were performed at the Vienna Scientific Cluster (VSC).
A. Weh thanks K. Held for the kind hospitality at TU Vienna. 
H. Schnait acknowledges financial support by the Austrian Science Fund (FWF), project No. Y746.
J. Otsuki was supported by JSPS KAKENHI Grant Nos. 18H01158 and 18H04301 (J-Physics).
Financial support offered by the Augsburg Center for Innovative Technologies,
and by the Deutsche Forschungsgemeinschaft (project number 107745057, TRR 80) is gratefully acknowledged.
\end{acknowledgments}


\appendix
\section{Padé analytic continuation}\label{sec:pade}
In the following, we show that a Padé analytic continuation of the Matsubara 
Green's function instead of the self-energy leads to artifacts in the spectrum.
The self-energy, \cref{eq:self} in \cref{subsec:bulk}, 
for \(T=\SI{0.25}{eV}\) can be fitted by the two pole function
\begin{equation}\label{eq:analytic_approx}
  \Sigma_\sigma (z) = \Sigma_{\sigma}^{HF} + \sum_{j=0}^{1} \frac{w_{\sigma j}}{z - \epsilon_{\sigma j}}
\end{equation}
with the residues \(w_{\sigma j}\) and poles \(\epsilon_{\sigma j}\) given in \cref{tab:self_parameters}.
\begin{table*}[ht]
  \begingroup\small
  \begin{ruledtabular}
  \begin{tabular}{lrrrr}
       & \multicolumn{1}{l}{\(w_{\sigma 0}\) (\si{eV^2})} & \multicolumn{1}{l}{\(\epsilon_{\sigma 0}\) (\si{eV})}
       & \multicolumn{1}{l}{\(w_{\sigma 1}\) (\si{eV^2})} & \multicolumn{1}{l}{\(\epsilon_{\sigma 1}\) (\si{eV})}
       \\\hline
       \(\sigma = {\uparrow}\) 
       & \num{+0.01066442-0.01755259j} & \num{-1.00707761-0.95609594j}
       & \num{+0.22879782+0.01755259j} & \num{+2.87718127-0.23207868j}
       \\
       \(\sigma = {\downarrow}\)
       & \num{-0.00282778-0.09524875j} & \num{+0.19827565-1.61153127j}
       & \num{+0.89041685+0.09524875j} & \num{+2.86888131-0.29918925j}
  \end{tabular}
  \end{ruledtabular}
  \endgroup
  \caption{%
      Residues \(w_{\sigma j}\) and poles \(\epsilon_{\sigma j}\) of \cref{eq:analytic_approx}
  }\label{tab:self_parameters}
\end{table*}
\begin{figure}[htb]
  \includegraphics[width=\linewidth]{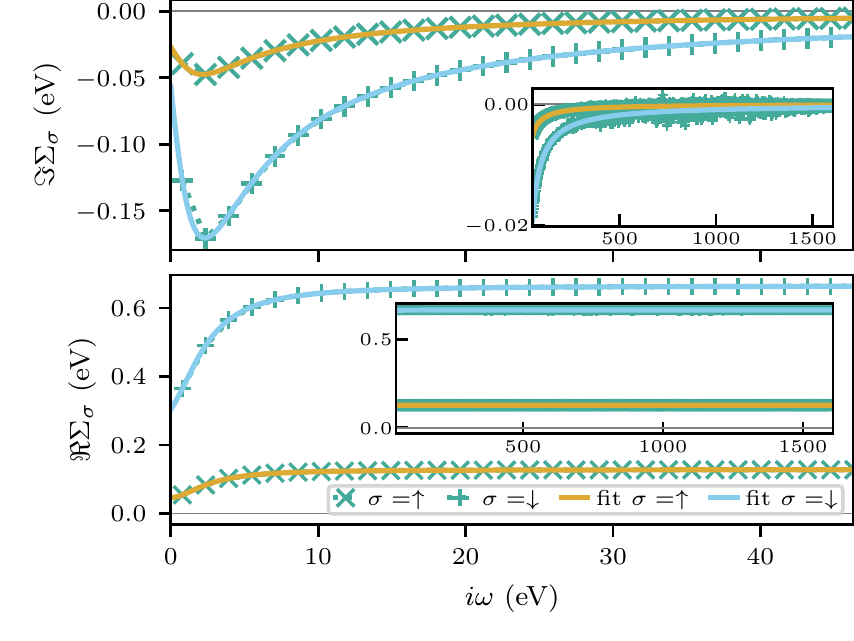}
  \vspace*{-0.5cm}
  \caption{%
    Imaginary (top) and real (bottom) part of the self-energy for \(T=\SI{0.25}{eV}\).
    The main plots show the data (markers) of the first \num{30} Matsubara frequencies,
    the continuous line is the fit \cref{eq:analytic_approx}.
    The insets show the remaining Matsubara frequencies (\num{31}-\num{1024});
    note the different scale for the imaginary part.
  }\label{fig:self-iw}
\end{figure}
Below, we will use this analytic expression as a realistic test case for the quality of analytic continuation.
First, we need to fit the parameters in \cref{eq:analytic_approx}.
They can be obtained using Padé analytic continuation of the self-energy \cref{eq:self},
as it yields an analytic formula (with numerical coefficients) in form of a rational polynomial
\(f(z) = p(z)/q(z)\), with polynomials \(p\) and \(q\).\cite{baker_essentials_1975}
The poles \(\epsilon_{\sigma j}\) can be calculated as the zeros of the denominator \(q\).
Instead of using Thiele's reciprocal difference method to determine the rational polynomial,
we directly calculate the poles in \cref{eq:analytic_approx} employing the algorithm 
presented in Ref.~\onlinecite{itoStablePolefindingRational2018}.
We write the linearized Padé approximation \(f(z)q(z) = p(z)\) in matrix form,
\begin{equation}
  \mat{F} \mat{V}_{q} \vec{q} = \mat{V}_{p} \vec{p} ,
\end{equation}
where \({(\mat{F})}_{ij} = f(z_{i}) \delta_{ij}\) is the diagonal matrix of function values,
\(\mat{V}_{q}\) and \(\mat{V}_{p}\) are the Vandermond matrices corresponding to \(q\) and \(p\),
and \(\vec{q}\) and \(\vec{p}\) are the polynomial coefficients.
We rewrite the equation as
\begin{equation}
  0
  = (\mat{F} \mat{V}_{q}, \mat{V}_{p}){(\vec{q}, -\vec{p})}^\mathrm{T}
  \eqqcolon \mat{C} \vec{x}.
\end{equation}
The number of poles \(M\) is then determined such
that the numerical null-dimension of the matrix \(\mat{C}\) is one.
To calculate the poles, we rewrite the rational polynomial \(f(z) = p(z)/q(z)\)
by factorizing a pole \(\epsilon_{m}\) from \(q(z) = (z - \epsilon_{m})\tilde{q}_{m}(z)\),
which leads to
\begin{equation}
  z f(z) \tilde{q}_{m}(z) - p(z) = \epsilon_{m} f(z) \tilde{q}_{m}(z).
\end{equation}
Again, we rewrite this set of equations in matrix form:
\begin{equation}
  (\mat{z F} \mat{V}_{\tilde{q}}, \mat{V}_{p}){(\vec{\tilde{q}}, - \vec{p})}^\mathrm{T}
  = \epsilon_{m} (\mat{F}\mat{V}_{\tilde{q}}, 0){(\vec{\tilde{q}}, - \vec{p})}^\mathrm{T}.
\end{equation}
This is a generalized eigenvalue problem, where the eigenvalues are the poles \(\epsilon_{m}\),
and the eigenvectors are the coefficients of the polynomials \(p\) and \(\tilde{q}\).
Using the knowledge of the poles \(\epsilon_{\sigma j}\),
the residues \(w_{\sigma j}\) can be obtained solving the linear equation in \(w_{\sigma j}\).
To reproduce noisy input,
Padé places artificial poles along the imaginary axis.
We verify that the residues of these poles are small and neglect them subsequently,
and recalculate the residues including physical poles only.\footnote{%
  It is necessary to verify that the contribution of unphysical poles is small.
  In case of a pole close to or on the real axis, like, e.g., in the Mott insulating phase,
  Padé might incorrectly place such poles in the upper complex plane.
  However, it is essential not to neglect these poles.
}

Using \cref{eq:analytic_approx}, we can evaluate the spectral function directly on the 
real axis
\begin{equation}\label{eq:analytic_spectrum}
\begin{aligned}
    A^{\text{analytic}}_\sigma(E)
    &= -\frac{1}{\pi} \Im\int\limits_{-D}^{D}\!\!\mathrm{d}E^{\prime}
    \frac{\rho(E^{\prime})}{E - E^{\prime} - \tilde{\epsilon}_{\sigma} - \Sigma_{\sigma}(E)}
    \\
    &= -\frac{1}{\pi} \Im g\big(E - \tilde{\epsilon}_{\sigma} - \Sigma_{\sigma}(E)\big),
\end{aligned}
\end{equation}
where the integral evaluates for the Bethe lattice with infinite coordination number to
\begin{equation}
  g(z)
  \coloneqq \int\limits_{-D}^{D}\!\!\mathrm{d}E \frac{\rho(E)}{z - E}
  = \frac{2z}{D^{2}} \left(1 - \sqrt{1 - {\left(\frac{z}{D} \right)}^{-2}}\right).
\end{equation}
We compare it with the Pad\'e analytic continuation \(A^{\text{Pad\'e}}_\sigma(E) = - \Im\widetilde{G}_{\sigma}(E)/\pi\) of the Green's function evaluated on the imaginary axis,
\begin{equation}\label{eq:gf_matsubara}
\begin{aligned}
  G_\sigma(\iw)
  &= \int\limits_{-D}^{D}\!\mathrm{d}E^{\prime}
  \frac{\rho(E^{\prime})}{\iw - E^{\prime} - \tilde{\epsilon}_{\sigma} - \Sigma_{\sigma}(\iw)}
  \\
  &= g\big(\iw - \tilde{\epsilon}_{\sigma} - \Sigma_{\sigma}(\iw)\big).
\end{aligned}
\end{equation}
In \cref{subfig:pade:self}, we use the fitted \(\Sigma_{\sigma}(z)\) as input to \cref{eq:gf_matsubara}.
\begin{figure}[htb]
  \begin{captivy}{\includegraphics[width=\linewidth]{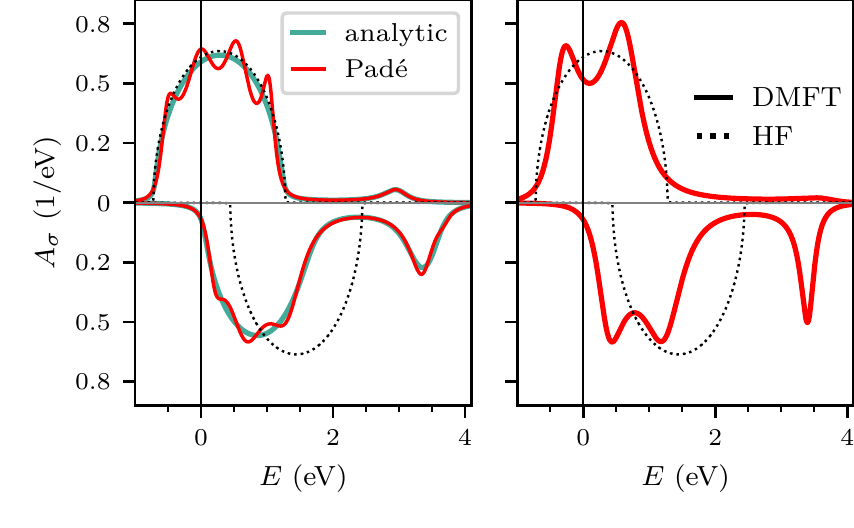}}
    \oversubcaption{0.19, 0.92}{}{subfig:pade:self}
    \oversubcaption{0.64, 0.92}{}{subfig:pade:gf}
  \end{captivy}
  \vspace*{-0.5cm}
  \caption{%
    \protect\subref{subfig:pade:self} Comparison of the Padé analytic continuation of the Matsubara Green's function,
    \cref{eq:gf_matsubara}, with the direct evaluation of \cref{eq:analytic_spectrum},
    using the fit \cref{eq:analytic_approx} as input to both.
    \protect\subref{subfig:pade:gf} Padé analytic continuation of the \emph{raw data} of the Matsubara Green's function
    \emph{not} employing the fit \cref{eq:analytic_approx}.
    The HF approximation is given as reference (dotted line).
  }\label{fig:pade}
\end{figure}
The Figure shows
that the analytic continuation of the Matsubara Green's function \cref{eq:gf_matsubara}
leads to spurious features similar to those in Ref.~\onlinecite{ch.ka.03},
in stark contrast to the analytic continuation of the self-energy used in \cref{fig:bulk}.
\Cref{subfig:pade:gf} shows the Padé continuation using the noisy raw data
\cref{eq:self} as input to \cref{eq:gf_matsubara}.
The result qualitatively agrees with the previous results\cite{ch.ka.03}.
We conclude that, independent of the presence of noise,
Pad\'e is unable to reproduce the branch-cut of the non-interacting Bethe DOS 
\(\rho(E)\) from Matsubara frequency data \cref{eq:gf_matsubara},
as it approximates the function with a finite number of poles.
The sharp band-edges cannot be resolved,
and oscillations similar to the Gibbs phenomenon in Fourier transform occur.
We can further support this argument by looking into the analytic continuation
of the hybridization function.
From \cref{eq:analytic_approx,eq:gf_matsubara} we can calculate the 
hybridization function on the Matsubara axis,
\begin{equation}
  \Delta_{\sigma}(\iw) = \iw - \tilde{\epsilon}_{\sigma} - \Sigma_{\sigma}(\iw) - 1/G_{\sigma}(\iw).
\end{equation}
This function encapsulates the effect of the DOS \(\rho(E)\);
if we perform the Padé analytic continuation \(\widetilde{\Delta}_{\sigma}(E)\),
we get the spectral function
\begin{equation}
  A^{\text{Padé hyb}}_{\sigma} (E) = - \frac{1}{\pi} \Im \frac{1}{E - \widetilde{\epsilon}_{\sigma} - \widetilde{\Delta}_{\sigma}(E) - \Sigma_{\sigma}(E)}
\end{equation}
which shows the same features as \(A^{\text{Padé}}_{\sigma}(E)=-\Im \widetilde{G}_{\sigma}(E)/ \pi\)
in \cref{fig:pade}.
An analytic continuation of the self-energy avoids this problem;
the analytic expression for \(\rho(E)\) is used directly,
and the self-energy lacks such sharp features.

\section{R-DMFT}\label{sec:RDMFT-derivation}
In the following, we derive the R-DMFT equations\cite{potthoff_metallic_1999,freericks_dynamical_2004}
used in \cref{subsec:hetero}.
We use an expansion in the coupling between the layers in the action formalism.
We start from the action \(S\) of the multilayer heterostructure,
which we split into two parts,
\begin{equation}\label{eq:S_heterostructure}
  S = \sum_{l} S_{l} + \Delta S.
\end{equation}
\(S_{l}\) is the action of the isolated layer \(l\),
and \(\Delta S\) contains the hopping in between the layers.
To ease the notation, we introduce the convention that indices with a bar are summed over:
\(\summed{l}\) is summed over layers,
\(\summed{\alpha}, \summed{\beta}\) over sites within a layer,
and \(\summed{\sigma}\) over spins.
The contributions to the action \cref{eq:S_heterostructure} reads
\begin{align}
  S_{l}
  &= \intdtau \left(%
    \gcc_{l \summed{\alpha}  \summed{\sigma}}
    [(\partial_{\tau} + \tilde{\epsilon}_{l}) \delta_{\summed{\alpha} \summed{\beta}} + t^{l}_{\summed{\alpha} \summed{\beta}}]
    \gc_{l \summed{\beta} \summed{\sigma}} 
  + U_{l} n_{l \summed{\alpha} \uparrow} n_{l \summed{\alpha} \downarrow}
  \right) 
  \\
  \Delta S
  &= \intdtau
  \gcc_{\summed{l} \summed{\alpha} \summed{\sigma}}
  t_{\summed{l}\summed{l^{\prime}}}
  \gc_{\summed{l^{\prime}} \summed{\alpha} \summed{\sigma}}.
\end{align}
We suppress the \(\tau\)-dependence of the Grassmann fields \(\gcc(\tau), \gc(\tau)\).
The action \(S_{l}\) and \(\Delta S\) correspond to the parts \(H_{l}\) and \( \hat{H}_{l l^{\prime}}\)
of the Hamiltonian, \cref{eq:ham_bilay}, neglecting the inter-layer Coulomb interaction.
The corresponding partition function reads
\begin{equation}\label{eq:Z_heterostructure}
  \Z = \intf{}[\gcc, \gc] \e^{-S[\gcc, \gc]}
  = \left[\prod_{l^{\prime}} \Z_{l^{\prime}}\right] \left\langle \e^{-\Delta S} \right\rangle_{\sum_{l} S_{l}}.
\end{equation}

Next we introduce auxiliary fields
and expand the contributions of the isolated layers in cumulants.
We truncate this expansion after the first order,
keeping only the quadratic part in the auxiliary fields.
This approximation results in a self-energy which is diagonal in the layers,
\(\Sigma_{l \alpha;l^{\prime} \beta} = \Sigma^{l}_{\alpha \beta} \delta_{ll^{\prime}}\).

We perform the \emph{Grassmannian Hubbard-Stratonovich} transformation following Ref.~\onlinecite{pairault_strongcoupling_2000}.
We rewrite the exponential \(\e^{- \Delta S}\) as Gaussian integral
over auxiliary Grassmann fields \(\aux_{l \alpha \sigma} (\tau)\) and \(\auxc_{l \alpha \sigma} (\tau)\):
\begin{equation}\label{eq:auxiliary-field_identity}
\begin{aligned}
  \e^{-\Delta S} 
  &= \frac{1}{\Zaux_{0}} \intf{}[\auxc, \aux]
  \e^{-\Saux_{0}
      - \intdtau%
      \left(%
        \auxc_{\summed{l} \summed{\alpha}  \summed{\sigma}} \gc_{\summed{l} \summed{\alpha} \summed{\sigma}}
        + \gcc_{\summed{l} \summed{\alpha} \summed{\sigma}} \aux_{\summed{l} \summed{\alpha} \summed{\sigma}}
      \right)
    },
\end{aligned}
\end{equation}
where we introduce the non-interacting auxiliary action
\begin{equation}\label{eq:def:S_aux}
  \Saux_{0}[\auxc, \aux]
  = -\intdtau \auxc_{\summed{l} \summed{\alpha} \summed{\sigma}}
  {(\mat{T}^{-1})}_{\summed{l} \summed{\alpha}; \summed{l^{\prime}} \summed{\beta}}
  \aux_{\summed{l} \summed{\beta} \summed{\sigma}},
\end{equation}
with the matrix \({(\mat{T})}_{l \alpha; l^{\prime} \beta} = t_{l l^{\prime}} \delta_{\alpha \beta}\).
The corresponding partition function \(\Zaux_{0}\) is
\begin{equation}
  \Zaux_{0} = \intf{}[\auxc, \aux] \e^{-\Saux_{0}}.
\end{equation}

Using \cref{eq:auxiliary-field_identity}, we rewrite the partition function \(\Z\)
as a field integral over auxiliary fields.
The average \(\langle \bcdot \rangle_{\sum_{l} S_{l}}\) in \cref{eq:Z_heterostructure}
can be split into the contributions of the isolated layers, yielding
\begin{multline}\label{eq:Z_of_auxillaries}
  \Z 
  = \frac{\prod_{l^{\prime}} \Z_{l^{\prime}}}{\Zaux_{0}} \intf{}[\auxc, \aux] 
    \e^{-\Saux_0 [\auxc, \aux]}
    \\
    \times\prod_{l} \left\langle \exp\Bigl(%
        -\intdtau (\auxc_{l \summed{\alpha} \summed{\sigma}} \gc_{l \summed{\alpha} \summed{\sigma}} + \gcc_{l \summed{\alpha} \summed{\sigma}} \aux_{l \summed{\alpha} \summed{\sigma}}
    )\Bigr) \right\rangle_{S_{l}}.
\end{multline}
We re-exponentiate the average \(\langle \bcdot \rangle_{S_{l}}\), and
expand the logarithm of the averages \(\langle \bcdot \rangle_{S_{l}}\) in terms of connected
Green's functions,
\begin{multline}\label{eq:auxiliary-field_cumulant-expansion}
  \ln\left\langle \exp\Bigl(%
    -\intdtau (\auxc_{l \summed{\alpha} \summed{\sigma}} \gc_{l \summed{\alpha} \summed{\sigma}} + \gcc_{l \summed{\alpha} \summed{\sigma}} \aux_{l \summed{\alpha} \summed{\sigma}}
 \Bigr) \right\rangle_{S_{l}}
 =
 \\
  -\sum_{n=1}^{\infty} {\sum_{\alpha_1 \dots \beta_n}}^{\!\!\!\prime} \intd \tau_{1} \dots \intd \tau_{n}^{\prime}
   \auxc_{l \beta_{1 \dots n}}
   \aux_{l \alpha_{1 \dots n}}
   G^{l}_{c\; \alpha_{1 \dots n}; \beta_{1 \dots n}},
\end{multline}
with the connected Green's function
\begin{equation}\label{eq:connected-green-function}
G^{l}_{c\;\alpha_{1 \dots n}; \beta_{1 \dots n}} =
-\langle \gc_{l \alpha_1} \dots \gc_{l \alpha_n} \gcc_{l \beta_1} \dots \gcc_{l \beta_n}\rangle_{S_l\;c} ;
\end{equation}
we use the abbreviation \(\auxc_{l \beta_{1 \dots n}} = \auxc_{l \beta_{1}} \dots \auxc_{l \beta_{n}}\),
and the prime at the sum denotes an ordered sum.
The \(\sigma\)-indices are suppressed.
The next step is to establish a relation between the fermionic Green's functions \(G\)
of the full lattice and the 
auxiliary field Green's functions \(\Gaux\).
To shorten the notation, we introduce new indices \(a\) and \(b\)
which denote sets \((l, \alpha, \sigma, \tau)\).
Evidently a fermionic \(n\)-particle Green's function can be generated by differentiating
the average \(\langle \bcdot \rangle_{\left(\sum_{l} S_{l}\right)}\) in \cref{eq:Z_of_auxillaries}
with respect to the auxiliary fields:
\begin{widetext}
\begin{equation}
\begin{aligned}
  G^{(n)}_{a_{1} \dots a_{n} b_{1} \dots b_{n}}
  &= - {\langle \gc_{a_{1}} \dots \gc_{a_n} \gcc_{b_1} \dots \gcc_{b_n} \rangle}_{S}
  \\
  &= -\frac{1}{\Z} \intf{}[\auxc, \aux] \e^{-\Saux_0}
   \frac{\delta^{2n}}{\delta \aux_{b_1} \dots \delta \aux_{b_n} \delta \auxc_{a_1} \dots \delta \auxc_{a_n}}
  \left\langle \e^{-\sum_{a} (\auxc_{a} \gc_{a} + \gcc_{a} \aux_{a})}\right\rangle_{\sum_{l} S_{l}}.
\end{aligned}
\end{equation}
Using integration by parts\cite{berezin_method_1966},
we can relate the fermionic \(G\) to the auxiliary fields Green's functions:
\begin{equation}
  G^{(n)}_{a_{1} \dots a_{n} b_{1} \dots b_{n}}
  = -\frac{1}{\Z} \intf{}[\auxc, \aux] \left(\e^{%
    -\Saux_0}
    \frac{{\stackrel{\leftharpoonup}{\delta}}^{2n}}{%
      \delta \aux_{b_1} \dots \delta \aux_{b_n} \delta \auxc_{a_1} \dots \delta \auxc_{a_n}
  }\right)
  \left\langle \e^{-\sum_{a} (\auxc_{a} \gc_{a} + \gcc_{a} \aux_{a})}\right\rangle_{\sum_{l} S_{l}}.
\end{equation}
\end{widetext}

The arrow in this equation indicates that the right derivative\cite{berezin_method_1966} is used, which means that the derivative acts from the right side on the Grassmann fields.
We explicitly calculate the expression for the one-particle Green's function.
The differentiation yields
\begin{multline}
  \e^{-\Saux_0} \frac{{\stackrel{\leftharpoonup}{\delta}}^{2}}{\delta \aux_{a} \delta \auxc_{b}} = 
  \\
  e^{-\Saux_{0}} \left[{(\mat{T}^{-1})}_{ab} + \sum_{a^{\prime} b^{\prime}} {(\mat{T}^{-1})}_{a b^{\prime}} \aux_{b^{\prime}} \auxc_{a^{\prime}} {(T^{-1})}_{a^{\prime} b}\right] ,
\end{multline}
and thus the one-particle Green's function is
\begin{equation}\label{eq:realation_G-G_aux}
  G_{ab} 
  \equiv G^{(1)}_{ab} 
  = -{(\mat{T}^{-1})}_{ba} + \sum_{a^{\prime} b^{\prime}} {(\mat{T}^{-1})}_{b a^{\prime}} \Gaux_{a^{\prime} b^{\prime}} {(\mat{T}^{-1})}_{b^{\prime} a}.
\end{equation}
We still need to calculate the auxiliary field Green's function \(\Gaux\).
At this point, we truncate the expansion \cref{eq:auxiliary-field_cumulant-expansion},
keeping only keep the first order, \(n=1\), hence
the action is quadratic in the auxiliary field:
\begin{equation}\label{eq:S_aux_first-order}
  \Saux 
  = \intdtau
  \auxc_{\summed{l} \summed{\alpha} \summed{\sigma}} \left(
    {-(\mat{T}^{-1})}_{\summed{l} \summed{\alpha}; \summed{l^{\prime}} \summed{\beta}}
    + \delta_{\summed{l} \summed{l^{\prime}}} G^{\summed{l}}_{\summed{\alpha} \summed{\beta}\, \summed{\sigma}} 
  \right) \aux_{\summed{l^{\prime}} \summed{\beta} \summed{\sigma}}.
\end{equation}
Then a Gaussian integration yields the auxiliary field Green's function \(\Gaux\):
\begin{equation}\label{eq:G_aux}
  \Gaux_{a b}
  = {\left(\mat{T}^{-1} - \diag (G^{l})\right)}^{-1}_{a b}.
\end{equation}
We plug this back into \cref{eq:realation_G-G_aux} and obtain the matrix equation
\begin{equation}\label{eq:G_heterostructure}
\begin{aligned}
  \mat{G} 
  &= - \mat{T}^{-1} + \mat{T}^{-1} {\left(\mat{T}^{-1} - \diag(G^{l})\right)}^{-1} \mat{T}^{-1} 
  \\
  &= {\left({\diag(G^{l})}^{-1} - \mat{T}\right)}^{-1}.
\end{aligned}
\end{equation}
Here the second equality is the Woodbury matrix identity\cite{higham_accuracy_2002}.
Thus we can calculate the Green's function \(G\) of the full heterostructure
from the Green's functions of the isolated layers \(G^{l}\).
Written in terms of the self-energy, we get
\begin{equation}\label{eq:G_heterostructure_realspace}
  \mat{G}^{-1} = \diag\big(\iw - \epsilon_{l} - \Sigma^{\prime}_{l}(\iw)\big) - \mat{T},
\end{equation}
where the self-energies \(\Sigma_{l}^{\prime}\) are determined by the Dyson equations of the isolated layers,
\begin{equation}\label{eq:Sigma_hetrostructure_approx}
  \Sigma^{\prime}_{l} = {G^{l}_{0}}^{-1} - {G^{l}}^{-1}.
\end{equation}
We determine these self-energies \(\Sigma_{l}^{\prime}\)
by using DMFT for every distinct isolated layer.
This approximation is evidently correct in the limit of isolated layers, \(\mat{T} = 0\),
and in the limit of non-interacting layers, \(U_{l} = 0\).
This approximation is supplemented by self-consistency.

\paragraph*{Self-consistent R-DMFT equations.}

We are going to replace the self-energy of the isolated layers, \(\Sigma^{\prime}_{l}\),
by a self-consistently determined self-energy.
We can also write \cref{eq:G_heterostructure_realspace} in the form
\begin{equation}
  \mat{G}^{-1} = \diag\left({(\mat{G}^{-1})}_{ll}\right) - \mat{T},
\end{equation}
i.e., the inverse Green's function is written as sum over the diagonal and 
the off-diagonal elements.
The Green's function then reads:
\begin{equation}\label{eq:G_heterostructure_self-consitent}
  \mat{G} 
  = {\left({\diag\left({(\mat{G}^{-1})}_{ll}\right)} - \mat{T}\right)}^{-1}.
\end{equation}
This looks like a self-consistency equation,
because it involves the same Green's function \(\mat{G}\) on the left- 
and right-hand-side of the equation.
Finally, the Dyson equation for the self-energy reads
\begin{equation}
  \Sigma_{l} = G_{0\, ll}^{-1} - G_{ll}^{-1}.
\end{equation}
We determine the self-energy \(\Sigma_{l}\) by a DMFT scheme\cite{potthoff_metallic_1999},
instead of using \(\Sigma^{\prime}_{l}\) from \cref{eq:Sigma_hetrostructure_approx}.

\bibliography{sources}

\end{document}

%% file: tikz/layer_structure.tikz.tex
\begin{tikzpicture}[xscale=1.5]
\colorlet{hopping_col}{green!70!black}
\tikzstyle{hopping}=[bend left, thick, hopping_col, latex-latex]
\pgfmathsetmacro{\N}{3}
\pgfmathsetmacro{\NN}{9}
\def\coordinationlist{1,...,\N}
\pgfmathsetmacro{\radius}{0.1}

\foreach \l in {2, 4}{
  \foreach \y in \coordinationlist{
    \foreach \z in \coordinationlist{
      \node[inner sep=\radius cm/1.6, circle] (\l_\y_\z) at (xyz cs:x=\l, y=\y, z=\z){};
      \filldraw[shading=ball] (\l_\y_\z) circle (\radius);
      \draw[hopping] (2_\y_\z) to ++(2, 0, 0);
    }
    \foreach \z [remember=\z as \zprev (initially 1)] in {2, ..., \N}{
      \draw[thick] (\l_\y_\zprev) to (\l_\y_\z);
    }
    \draw[thick, dashed] (\l_\y_1) to ++(0, 0, -0.5);
    \draw[thick, dashed] (\l_\y_\N) to ++(0, 0, +0.5);
  }
  \foreach \z in \coordinationlist{
    \foreach \y [remember=\y as \yprev (initially 1)] in {2, ..., \N}{
      \draw[thick] (\l_\yprev_\z) to (\l_\y_\z);
    }
    \draw[thick, dashed] (\l_1_\z) to ++(0, -0.5, 0);
    \draw[thick, dashed] (\l_\N_\z) to ++(0, +0.5, 0);
  }
  \draw[fill=gray, opacity=0.4] (\l_1_1)++(0, -1, -1) to ++(0, 0, \N+1) to ++ (0, \N+1, 0) to ++ (0, 0, -\N-1) to cycle;
}
\node at (2.5, 3.50,0){\Large \textcolor{hopping_col}{\(t\)}};

\end{tikzpicture}